\documentclass[a4paper,11pt]{article}
\usepackage{jheppub}
\usepackage[dvipsnames,table]{xcolor}
\usepackage{xspace,adjustbox}
\usepackage[tight]{subfigure}
\usepackage{amsmath}
\usepackage{amssymb}
\usepackage{amsfonts}
\usepackage[title,titletoc]{appendix}
\usepackage{tikz}
\usepackage[normalem]{ulem}
\usepackage{physics}
\usepackage{cleveref}
\usepackage{multirow}
\usepackage{graphicx}

\def\refeq#1{eq.~(\ref{#1})}
\def\reffi#1{\mbox{figure~\ref{#1}}}
\def\reffis#1{\mbox{figures~\ref{#1}}}
\def\refta#1{\mbox{table~\ref{#1}}}

\def\refse#1{\mbox{section~\ref{#1}}}
\def\refapp#1{\mbox{appendix~\ref{#1}}}
\def\citere#1{\mbox{ref.~\cite{#1}}}

\def\Refse#1{\mbox{Section~\ref{#1}}}

\newcommand{\GeV}{\ensuremath{\,\text{GeV}}\xspace}
\newcommand{\TeV}{\ensuremath{\,\text{TeV}}\xspace}
\newcommand{\fb}{{\ensuremath\unskip\,\text{fb}}\xspace}

\newcommand{\eg}{\emph{e.g.}\ }

\newcommand{\eps}{\varepsilon}

\newcommand{\ri}{\mathrm i}

\newcommand{\rV}{{\mathrm{V}}}
\newcommand{\rT}{{\mathrm{T}}}

\newcommand{\rL}{{\mathrm{L}}}
\newcommand{\rF}{{\mathrm{F}}}

\newcommand{\Pj}{\ensuremath{\text{j}}\xspace}
\newcommand{\Pp}{\ensuremath{\text{p}}}
\newcommand{\Pe}{\ensuremath{\text{e}}\xspace}

\newcommand{\Pq}{\ensuremath{q}}

\newcommand{\Pu}{\ensuremath{\text{u}}\xspace}
\newcommand{\Pd}{\ensuremath{\text{d}}\xspace}

\newcommand{\Pc}{\ensuremath{\text{c}}\xspace}
\newcommand{\Pg}{\ensuremath{\text{g}}}

\newcommand{\PW}{\ensuremath{\text{W}}\xspace}
\newcommand{\PZ}{\ensuremath{\text{Z}}\xspace}

\newcommand{\MWOS}{\ensuremath{M_\PW^\text{OS}}\xspace}
\newcommand{\MW}{\ensuremath{M_\PW}\xspace}
\newcommand{\MV}{\ensuremath{M_\rV}\xspace}
\newcommand{\MZOS}{\ensuremath{M_\PZ^\text{OS}}\xspace}

\newcommand{\MVOS}{\ensuremath{M_{V}^\text{OS}}\xspace}%
\newcommand{\GVOS}{\ensuremath{\Gamma_{V}^\text{OS}}\xspace}%

\newcommand{\GZOS}{\ensuremath{\Gamma_\PZ^\text{OS}}\xspace}

\newcommand{\GWOS}{\ensuremath{\Gamma_\PW^\text{OS}}\xspace}

\newcommand{\alphas}{\ensuremath{\alpha_\text{s}}\xspace}

\newcommand{\Wplusj}{\PW\!+\Pj}
\newcommand{\Wplusc}{\PW\!+\!\Pc}
\newcommand{\ppmnj}{\Pp\Pp\rightarrow\ell^\pm\,\overset{\scriptscriptstyle(-)}{\nu}_{\!\!\!\ell}\,\Pj}
\newcommand{\ppmnjM}{\Pp\Pp\rightarrow\ell^-\,\bar\nu_\ell \Pj}
\newcommand{\ppmnjP}{\Pp\Pp\rightarrow\ell^+\,\nu_\ell \Pj}

\newcommand{\ptj}{p_{\rT,\Pj}}

\newcommand{\AvH}{{\sc AvH}\xspace}
\newcommand{\cpp}{{\sc C++}\xspace}

\newcommand{\Stripper}{{\sc Stripper}\xspace}

\newcommand{\OpenLoops}{O\protect\scalebox{0.8}{PEN}L\protect\scalebox{0.8}{OOPS}\xspace}

\newcolumntype{.}{D{.}{.}{-1}}
\newcolumntype{d}[1]{D{.}{.}{#1}}
\newcommand{\ordercoupling}[1]{\ensuremath{\mathcal{O}{\left(#1\right)}}\xspace}

\newcolumntype{C}[1]{>{\centering\arraybackslash}p{#1}}

\dedicated{\rm {CAVENDISH-HEP-21/13}, {FR-PHENO-2021-11}}
\title{Polarised W+j production at the LHC: \\
a study at NNLO QCD accuracy}

\author{Mathieu Pellen$^1$,}
\author{Rene Poncelet$^2$,}
\author{and Andrei Popescu$^2$}

\affiliation{$^1$Universit\"at Freiburg,
        Physikalisches Institut, \\
        Hermann-Herder-Stra\ss e 3,
        79104 Freiburg,
        Germany}
\affiliation{$^2$Cavendish Laboratory, University of Cambridge,\\
             J.J. Thomson Avenue, Cambridge CB3 0HE, United Kingdom}

\emailAdd{mathieu.pellen@physik.uni-freiburg.de}
\emailAdd{poncelet@hep.phy.cam.ac.uk}
\emailAdd{popescu@hep.phy.cam.ac.uk}
\date{\draftdate}

\abstract
{
  We study polarisation of W-bosons produced in
  association with one jet at the LHC. In particular, we provide all necessary
  theoretical ingredients for the precise extraction of polarisation fractions.
  To that end, we present new polarised predictions up to NNLO QCD accuracy
  employing the narrow-width approximation, in two phase spaces: inclusive and fiducial.
  We compare results in the fiducial phase space to a full off-shell
  computation as well as experimental data. Finally, we fit the polarisation
  fractions using shape templates and show that NNLO corrections significantly
  improve their determination.
}


\begin{document}

\maketitle

\section{Introduction}\label{sec:introduction}

High-energy collisions of protons at the Large Hadron Collider (LHC) allow for
a copious production of W-bosons in association with one or more high-$p_{\rm
T}$ jets. In fact, this class of processes has one of the highest cross sections for
electroweak (EW) boson production and therefore constitutes an important
background to many other Standard Model processes. This motivated numerous
theoretical and experimental works \cite{Azzurri:2020lbf}.

By scrutinising the longitudinal mode, which originates from the electroweak
symmetry breaking (EWSB) mechanism, one gets an insight into the EW
sector with possible footprints of new-physics effects.
This idea has lead to numerous massive vector boson polarisation studies for
different LHC processes, including Drell-Yan \cite{Manca:2017xym},
diboson production
\cite{Baglio:2018rcu,Baglio:2019nmc,Denner:2020bcz,
Denner:2020eck,Denner:2021csi},
and vector-boson scattering
\cite{Ballestrero:2017bxn,Ballestrero:2019qoy,Ballestrero:2020qgv}.
The production of a polarised W-boson in association with one jet
represents another important handle to pinpoint the SM, and it has therefore
received its due attention in the scientific community
\cite{Bern:2011ie,Stirling:2012zt,Belyaev:2013nla},
and several dedicated experimental measurements \cite{ATLAS:2012au,CMS:2011kaj}
were carried out.

From a theoretical point of view, $\Wplusj$ production is a rather well
understood process, with NNLO QCD predictions
\cite{Boughezal:2015dva,Boughezal:2016dtm,Gehrmann-DeRidder:2019avi}, EW
corrections \cite{Kuhn:2007qc,Kuhn:2007cv,Hollik:2007sq,Denner:2009gj},
and combinations thereof
\cite{Kallweit:2014xda,Kallweit:2015dum,Biedermann:2017yoi,Lindert:2017olm}
being already available.
Nonetheless, predictions for the polarised W-bosons have so far been limited to
NLO QCD accuracy \cite{Bern:2011ie,Stirling:2012zt}.

In this work, we are filling this gap by providing, for the first time, the NNLO QCD
predictions for the polarised W-bosons in the narrow-width approximation
(NWA) at the LHC running at 13 \TeV. We perform this computation in two
different phase spaces: \emph{inclusive} and \emph{fiducial}.
The inclusive phase space gives a clean and a more holistic picture of polarisation
features, whereas the fiducial phase space mimics a realistic measurement and
helps understand the limitations of the NWA and boson-polarisation definition.
Thus, in the fiducial phase space, we also compare the obtained NWA predictions with
the full off-shell computation as well as with experimental data provided by
the CMS collaboration \cite{CMS:2017gbl}.
The first comparison provides a realistic quality check of the NWA,
which plays a crucial role in the definition of the polarised states.
Finally, we use polarised distributions, obtained
at amplitude level, as templates to fit the polarisation fractions to the data,
and show that NNLO corrections significantly improve their precise determination.

The article is organised as follows: in \refse{sec:setup} we introduce
the process under consideration and goes over details of our computation.
We list numerical inputs used as well as technical details for both phase spaces.
We provide information on the implementation and list validations
that we have carried out.
\Refse{sec:results} presents results of the computation. This
includes: theoretical predictions for the polarised W-boson in the form of cross
sections and differential distributions, polarisation fractions, comparison to
off-shell simulations and data, and finally, fits that can be used to extract
polarisation fractions. We summarise our findings and make concluding remarks in
\refse{sec:conclusions}.

\section{Setup}\label{sec:setup}

\subsection{Definition of the process}
\label{sec:def}

\begin{figure}
    \center
      {\includegraphics[width=.39\textwidth]{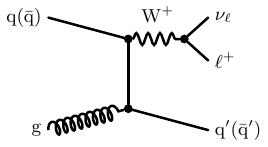}}
      \qquad\quad
      {\includegraphics[width=.39\textwidth]{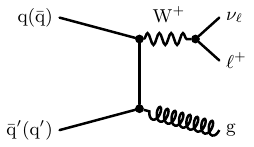}}
      \caption{
        Leading-order Feynman diagrams of the hadronic process $\ppmnjP$.
        \label{fig:diag}%
      }
\end{figure}

In this work, the hadronic processes under investigation are
\begin{equation}
  \ppmnj\;,
\end{equation}
and are usually referred to as $\Wplusj$ production.
In this study, given that we consider massless leptons, both cases $\ell=\Pe,\mu$
are equivalent. In \reffi{fig:diag}, two
LO diagrams for the ``plus'' signature are represented. The LO is defined at
order $\ordercoupling{\alpha^2\alphas}$, while the NLO and NNLO QCD corrections
are defined at orders $\ordercoupling{\alpha^2\alphas^2}$ and
$\ordercoupling{\alpha^2\alphas^3}$, respectively. Further on, we will
use the following notation for the cross sections and their perturbative
expansions:
\begin{align}
  \sigma(\ppmnj) &\equiv \sigma^{\pm} = \sigma^{\pm(0)}
                                       +\sigma^{\pm(1)}
                                       +\sigma^{\pm(2)}
                                       +\order{\alphas^4}\,.
\end{align}
Also, we consider both the ``plus'' and ``minus'' signatures
separately unless explicitly stated otherwise, so we will omit the $\pm$
superscript for brevity. Finally, we define
\begin{align}
  \sigma^{\pm}_{\rm LO}   &= \sigma^{\pm(0)}\,,\nonumber\\
  \sigma^{\pm}_{\rm NLO}  &= \sigma^{\pm(0)}+\sigma^{\pm(1)}\nonumber\,, \\
  \sigma^{\pm}_{\rm NNLO} &= \sigma^{\pm(0)}+\sigma^{\pm(1)}+\sigma^{\pm(2)}\,,
\end{align}
and the corresponding $K$-factors to express perturbative corrections as
\begin{align}
  K^{\pm}_{\rm NLO} = \frac{\sigma^{\pm}_{\rm NLO}}{\sigma^{\pm}_{\rm LO}}\,,
  \quad K^{\pm}_{\rm NNLO} =
    \frac{\sigma^{\pm}_{\rm NNLO}}{\sigma^{\pm}_{\rm NLO}}\,.
\end{align}
The same notation will be used analogously in differential cross sections.
$K$-factors are generally given with respect to the central scale choice,
and scale variations are given by variations in the numerator only.

In addition, to QCD corrections, the EW ones are also phenomenologically relevant.
For $\Wplusj$ production at the LHC, they are typically at the level
of a few per cent for total cross sections but can reach several
dozen per cent in the high energy limits \cite{Denner:2009gj}.
For $\Wplusj$ polarised predictions, EW corrections are unknown\footnote{At
the moment, NLO EW corrections for polarised processes
are only available for $\PZ\PZ$ production at the LHC \cite{Denner:2021csi}.}.
The study of the impact of EW corrections for $\Wplusj$ polarised predictions
is thus left for future work.
In this work, we focus exclusively on the effect of QCD corrections.

In order to define the cross sections for polarised W-bosons, we need to define
polarisation states of the propagating bosons.
This requires the bosons to be on their mass shell.
To this end, we employ the NWA which
allows to factorise the production and decay of on-shell bosons, while keeping
all spin information. For the process at hand, this approximation does not lead
to a reduction of the number of Feynman diagrams to be considered. Indeed, in
the present case, the leptonic pair consisting of a neutrino and a charged
lepton can only be obtained through decay of the W-boson, in contrast to
processes involving neutral currents. Considering the NWA, we write tree-level
amplitudes, as well as virtual and higher-multiplicities
amplitudes, in the following way:
\begin{align}
 \mathcal{M}^{\rm NWA}(\ppmnj) =
    \frac{1}{M_W\Gamma_W}\sum_{h\in \Lambda} \mathcal{M}_h (\Pp\Pp \to \PW \Pj)\cdot
    \Gamma_h\left(\PW \to \ell^\pm\,\overset{\scriptscriptstyle(-)}{\nu}_{\!\!\!\ell}\right),
\end{align}
where $h$ represents polarisations from the set $\Lambda = \{+1,-1,0\}$.
The `transversely polarised' amplitudes is represented by $\Lambda = \rT = \{+1,-1\}$
and the `longitudinally polarised' by $\Lambda = \rL = \{0\}$.
By squaring this amplitude, one obtains unpolarised cross sections.
In turn, polarised cross sections are retrieved by squaring only one polarised
amplitude. It implies that the sum of the polarised cross sections and the
unpolarised ones differ
by the interference contributions between the different polarised amplitudes.
Such effects are analysed in details in section~\ref{sec:OS_IE_data}.

The definition of boson polarisation is not unique. To be
exact, with W-boson momentum in the laboratory frame
$p = (E,\vec{p}) = (E,
|\vec{p}|\sin\theta\cos\phi, |\vec{p}|\sin\theta\sin\phi, |\vec{p}|\cos\theta)$,
we define the polarisation vector using the helicity frame
\cite{Bern:2011ie,Poncelet:2021jmj} as follows:
\begin{align}
  \eps^{\mu}_{-} &= \frac{1}{\sqrt 2}
                    \left(0,\cos\theta\cos\phi +  \ri\sin\phi,
                    \cos\theta\sin\phi - \ri\cos\phi, - \sin\theta \right), \notag \\
  \eps^{\mu}_{+} &= \frac{1}{\sqrt 2}
                    \left(0,- \cos\theta\cos\phi +  \ri\sin\phi ,
                     - \cos\theta\sin\phi - \ri\cos\phi, \sin\theta\right), \notag \\
  \eps^{\mu}_{0} &= \frac{1}{m_W}
                    \left(|\vec{p}|, E\sin\theta\cos\phi, E\sin\theta\sin\phi,
                    E\cos\theta\right).
\end{align}

To quantify the contribution of a particular polarisation to the (differential)
cross section we use polarisation fractions
\begin{align}
  f_p = \frac{\sigma_p}{\sum_p \sigma_p} \quad \text{for}\quad p \in \{\rm L,T\}\,,
\end{align}
where $\sigma_p$ is the cross section with polarisation $p$. For estimating
the scale uncertainty in the ratio, we use an uncorrelated estimation, that is
we perform the squared error propagation of independent scale variations in both
the numerator and denominator.

\subsection{Numerical inputs and event selections}

\subsection*{Numerical inputs}

Our computation simulates $\ppmnj$ processes at the LHC
running with the centre-of-mass energy of $\sqrt{s}=13\TeV$. The parton
distribution functions (PDF) used are the \texttt{NNPDF31\_as\_0118} set
\cite{NNPDF:2017mvq} with corresponding orders.
Throughout the computation we use the $n_f = 5$ scheme.
The input masses and widths are
\begin{alignat}{2}
  \MWOS &= 80.3790 \GeV, \qquad &
  \GWOS &= 2.0850  \GeV, \\
  \MZOS &= 91.1876 \GeV, \qquad &
  \GZOS &= 2.4952  \GeV.
\end{alignat}
They determine the \emph{pole} boson parameters actually used in the numerical
evaluation of the matrix elements \cite{Bardin:1988xt}:
\begin{equation}
  \MV = \frac{\MVOS}{\sqrt{1+(\GVOS/\MVOS)^2}},
  \qquad \Gamma_\rV = \frac{\GVOS}{\sqrt{1+(\GVOS/\MVOS)^2}},
\end{equation}
for $\rV = \PW,\PZ$. The electromagnetic coupling is fixed
following the $G_\mu$ scheme \cite{Denner:2000bj} with
\begin{equation}
  G_\rF = 1.16638 \times 10^{-5} \GeV^{-2}.
\end{equation}
All leptons are considered massless and we assume the diagonal CKM matrix.

The central choice used for both factorisation and renormalisation scale reads
\begin{equation}
\label{eq:mu}
  \mu = \frac12 H_\rT'
  = \frac12 \left( \sqrt{\MW^2 + p^2_{\rT,\PW}}
  + \sum_{i=\textrm{jets}} p_{\rT,i}\right),
\end{equation}
where the sum goes over all jets obtained after applying the jet algorithm.
For the estimate of the missing higher orders, we adopt the standard
7-point scale variation prescription.

\subsection*{Phase-space definitions}

In the present work, we consider two phase spaces that we refer to as
\emph{inclusive} and \emph{fiducial} setup, respectively.

In both cases, we use the anti-$k_\text{T}$ algorithm \cite{Cacciari:2008gp} with
$R=0.4$ to identify QCD jets. We require at least one jet fulfilling
\begin{equation}
 \abs{y_\Pj} < 2.4 \quad \textrm{and} \quad \ptj>30\GeV.
 \label{eq:jet}
\end{equation}

\begin{itemize}
  \item \underline{Inclusive setup:} \\
    The inclusive setup does not have any cuts apart from the jet acceptance
    criteria in \refeq{eq:jet}.
  \item \underline{Fiducial setup:} \\
    To emulate a realistic experimental setup, we follow the selection procedure
    used in the recent CMS study \cite{CMS:2017gbl}.
    In addition to \refeq{eq:jet}, jets are required to have
    \begin{equation}
      \Delta R(\ell,j) > 0.4,
      \label{eq:deltaR}
    \end{equation}
    where $\ell$ is the charged lepton originating from the boson decay.
    Furthermore, the following requirements are applied to the leptonic products:
    \begin{align}
      p_{{\rT},\ell} > 25\GeV, \qquad |\eta_\ell| < 2.5, \qquad M_{\rT}(\PW) > 50\GeV ,
    \end{align}
    where the W-boson transverse mass is defined as
    \begin{equation}
      m_{\rT}^{\PW} = \sqrt{M_\PW^2 + p_{\rT,\PW}^2}
                    = \sqrt{2 p_{{\rT},\ell} \cdot p_{\rT,\text{miss}}
                      \left(1 - \cos\Delta\phi\right)},
    \label{eq:mTW}
    \end{equation}
    with $\Delta \phi = \min \left( |\phi_\ell - \phi_\nu|,
                             2\pi - |\phi_\ell - \phi_\nu| \right)$
    being the azimuthal separation of the lepton and neutrino momenta.
    We define $p_{\rT,\text{miss}}$ using the neutrino momentum.
\end{itemize}

\subsection{Implementation and validation procedures}

The computation has been carried out using the \Stripper framework which is a
\cpp implementation of the four-dimensional formulation of the sector-improved
residue subtraction scheme \cite{Czakon:2010td,Czakon:2014oma,Czakon:2019tmo}.
The \Stripper library consists in a Monte-Carlo generator which automates the
subtraction procedure. For the computation of matrix elements, it relies on
external programs. Tree-level amplitudes, are calculated using \AvH library
\cite{Bury:2015dla}. The one-loop amplitudes are calculated using our
privately extended version of \OpenLoops 2 \cite{Cascioli:2011va,Buccioni:2017yxi,
Buccioni:2019sur}, which is
capable of calculations with defined polarisations. The two-loop helicity
amplitudes for the off-shell calculation and the necessary components for the
polarised studies were extracted from \citere{Gehrmann:2011ab}. The adaptation
of these amplitudes to on-shell polarised bosons can be found in the
\refapp{sec:appendix}.

The validity of this calculation is supported by the following considerations
and cross checks.
The calculation exploits the same framework that was used for the NNLO~QCD
study of $\Wplusc$ \cite{Czakon:2020coa} and the polarised $\PW^+\PW^-$ production
\cite{Poncelet:2021jmj}. The latter study also
explored the effectiveness of the NWA approach in the polarisation setting.
Details about the implementation of the NWA can be found in ref.~\cite{Czakon:2020qbd}.
All 0-, 1-, and 2-loop on-shell amplitudes were checked against full off-shell
amplitudes at on-shell phase space points.
The 0- and 1- loop polarised amplitudes were also cross-checked between \AvH,
\OpenLoops 2, {\sc Recola} \cite{Actis:2012qn,Actis:2016mpe}, and \citere{Gehrmann:2011ab}.
Finally, we produced polarised predictions
at NLO following the setup of \citere{Bern:2011ie} and found perfect agreement
with their results.

\section{Results}\label{sec:results}

In this section we present and analyse results of our numerical simulations. At
first, we discuss polarisation setups in \refse{subsec:polarised_predictions}.
Then we explore the differences between the ``plus'' and ``minus'' signature in
section \ref{sec:charges}. Further on, we discuss off-shell and interference
effects and compare our predictions against experimental data of the CMS
collaboration \cite{CMS:2017gbl}. Finally, we perform a fit of these
experimental data using the polarised shape distributions we have computed in
the previous parts.

\subsection{Polarised predictions}\label{subsec:polarised_predictions}

As explained in \refse{sec:setup}, once a W-boson is on its mass shell,
it is possible to define its longitudinal (L) and
transverse (T) polarisations.
The total cross sections of such polarised predictions
for both the inclusive and fiducial setups are presented in
tables~\ref{tab:inclusive_xsec} and \ref{tab:fiducial_xsec}, respectively.

\begin{table}[!h]
  \centering
  {\small
  \renewcommand{\arraystretch}{1.3}
\begin{tabular}{|C{1.0cm}|C{2.26cm}C{2.26cm}C{1.0cm}C{2.26cm}C{1.0cm}C{2.2cm}|}
    \hline
        \cellcolor{blue!9}{\textit{Inc.}}
      & \cellcolor{blue!9}{LO [fb]}
      & \cellcolor{blue!9}{NLO [fb]}
      & \cellcolor{blue!9}{$K_\text{NLO}$}
      & \cellcolor{blue!9}{NNLO [fb]}
      & \cellcolor{blue!9}{$K_\text{NNLO}$}
      & \cellcolor{blue!9}{$f_\text{NNLO}$}
      \\
    \hline
    $\PW^+_{\rm L}$
        & $162.964(8)^{+11.5\%}_{-9.4\%}$
        & $268.4(2)^{+9.4\%}_{-8.0\%}$
        & 1.65
        & $287.8(10)^{+1.3\%}_{-2.8\%}$
        & 1.07
        & $0.186(2)^{+3.2\%}_{-3.1\%}$
    \\
    $\PW^+_{\rm T}$
        & $827.00(7)^{+11.4\%}_{-9.3\%}$
        & $1214.5(5)^{+7.6\%}_{-6.8\%}$
        & 1.47
        & $1258(6)^{+0.3\%}_{-1.7\%}$
        & 1.04
        & $0.814(11)^{+2.2\%}_{-2.0\%}$
    \\
    \hline
    $\PW^-_{\rm L}$
        & $133.386(6)^{+11.5\%}_{-9.4\%}$
        & $208.2(1)^{+8.4\%}_{-7.4\%}$
        & 1.56
        & $220.5(7)^{+1.0\%}_{-2.4\%}$
        & 1.06
        & $0.191(2)^{+2.6\%}_{-2.5\%}$
     \\
    $\PW^-_{\rm T}$
        & $640.58(5)^{+11.5\%}_{-9.3\%}$
        & $913.0(3)^{+6.8\%}_{-6.3\%}$
        & 1.43
        & $934(3)^{+0.2\%}_{-1.3\%}$
        & 1.02
        & $0.809(9)^{+1.7\%}_{-1.5\%}$
    \\
    \hline
  \end{tabular}
  }
  \caption{Total cross sections of $\ppmnj$ for the \emph{inclusive} setup at the LHC
    with the centre-of-mass energy of $\sqrt{s}=13\TeV$. The polarised predictions are
    provided at LO, NLO, and NNLO accuracy along with the corresponding
    $K$-factors. The last column represents polarisation fractions at NNLO, calculated
    as a ratio of the polarised result to the sum of the two polarisations. The numbers in
    parentheses indicate statistical errors in the last significant digit,
    while sub- and superscripts
    represent scale variation uncertainty with respect to the full result.
  }
  \label{tab:inclusive_xsec}
\end{table}

\begin{table}[!h]
  \centering
  {\small
  \renewcommand{\arraystretch}{1.3}
\begin{tabular}{|C{1.0cm}|C{2.26cm}C{2.26cm}C{1.0cm}C{2.26cm}C{1.0cm}C{2.2cm}|}
    \hline
        \cellcolor{blue!9}{\textit{Fid.}}
      & \cellcolor{blue!9}{LO [fb]}
      & \cellcolor{blue!9}{NLO [fb]}
      & \cellcolor{blue!9}{$K_\text{NLO}$}
      & \cellcolor{blue!9}{NNLO [fb]}
      & \cellcolor{blue!9}{$K_\text{NNLO}$}
      & \cellcolor{blue!9}{$f_\text{NNLO}$}
      \\
    \hline
    $\PW^+_{\rm L}$
        & $93.898(5)^{+11.5\%}_{-9.4\%}$
        & $147.07(4)^{+8.3\%}_{-7.3\%}$
        & 1.57
        & $156(1)^{+1.4\%}_{-2.7\%}$
        & 1.06
        & $0.246(6)^{+3.1\%}_{-2.9\%}$
    \\
    $\PW^+_{\rm T}$
        & $319.31(3)^{+11.5\%}_{-9.3\%}$
        & $466.5(1)^{+6.7\%}_{-6.2\%}$
        & 1.46
        & $477(2)^{+0.2\%}_{-1.3\%}$
        & 1.02
        & $0.754(9)^{+1.9\%}_{-1.5\%}$
    \\
    \hline
    $\PW^-_{\rm L}$
        & $79.831(5)^{+11.5\%}_{-9.4\%}$
        & $122.56(7)^{+8.0\%}_{-7.1\%}$
        & 1.54
        & $128.8(4)^{+0.7\%}_{-2.1\%}$
        & 1.05
        & $0.245(3)^{+2.1\%}_{-2.3\%}$
    \\
    $\PW^-_{\rm T}$
        & $273.08(2)^{+11.5\%}_{-9.3\%}$
        & $389.8(1)^{+6.2\%}_{-5.9\%}$
        & 1.43
        & $396(1)^{+0.2\%}_{-1.1\%}$
        & 1.02
        & $0.755(6)^{+1.5\%}_{-1.3\%}$
   \\
   \hline
  \end{tabular}
  }
  \caption{Total cross sections of $\ppmnj$ for the \emph{fiducial} setup at the LHC
    with $\sqrt{s}=13\TeV$. This table has the same layout
    as \refta{tab:inclusive_xsec}.
  \label{tab:fiducial_xsec}
  }
\end{table}

Firstly, the cross sections for the ``plus''
and ``minus'' signatures are rather different. This is due to the fact that
they have different PDF contributions: the ``plus'' signature is dominated by
$\Pu\Pg$-channel, and the ``minus'' one --- by $\Pd\Pg$-channel. Secondly, the
NLO $K$-factors are significantly larger than the NNLO ones. This is by now a
relatively well understood phenomenon which is originates from the new topologies
appearing at NLO \cite{Rubin:2010xp}. It affects both signatures, which is
reflected in the similarity of their QCD corrections.
Thirdly, the scale variation is brought
down by a factor of 5 by NNLO corrections. Combined with the fact that the
$K$-factor is small at NNLO, this is a great sign for convergence of the QCD
perturbative series. Finally, the polarisation fractions differ in the fiducial
and inclusive setups, 
albeit having similar NNLO corrections in both cases.
\Refse{sec:charges} is dedicated to explore
differences between the two signatures in more detail.

\begin{figure}
  \centering
  \includegraphics[width=.49\linewidth]{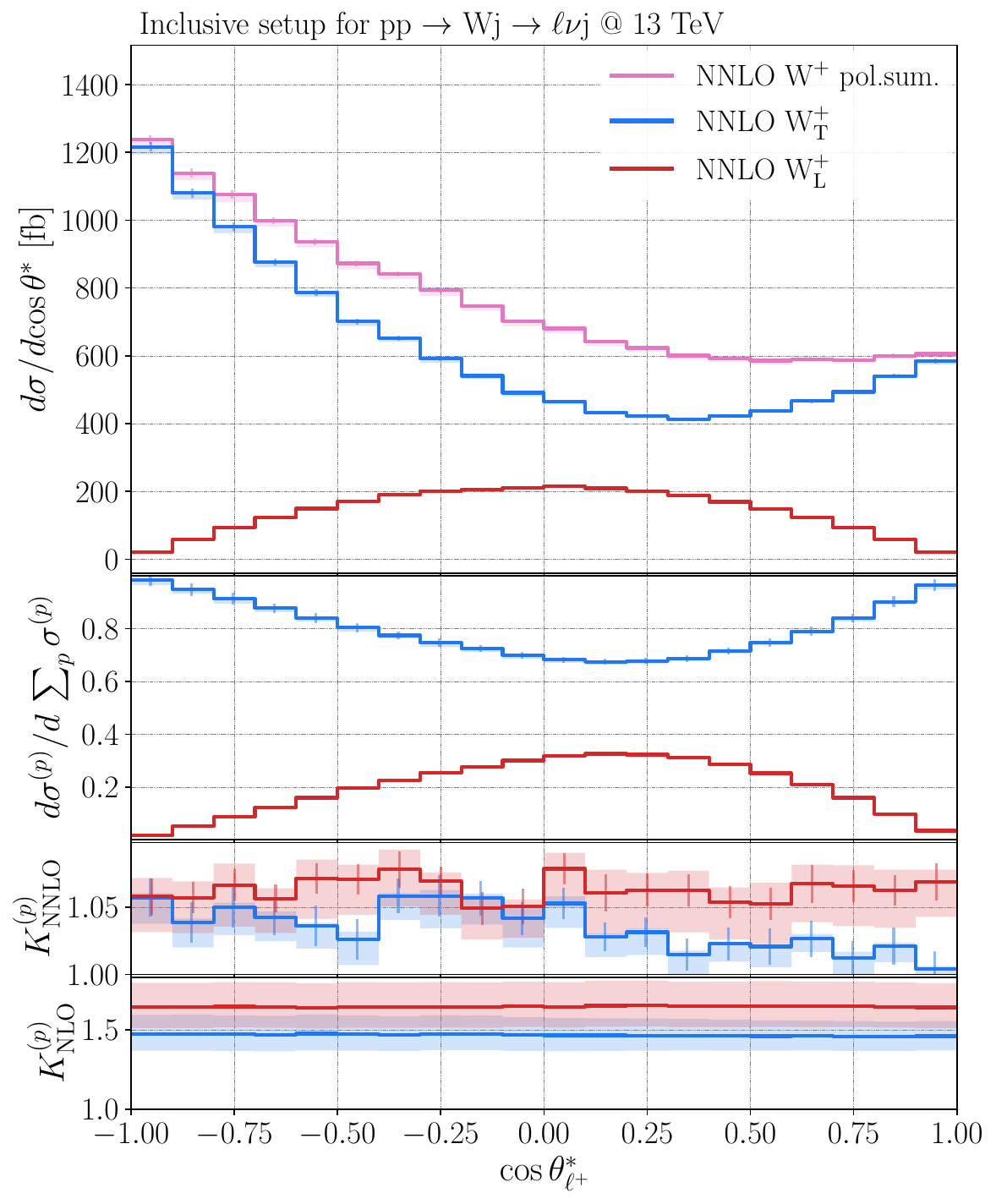}
  \includegraphics[width=.49\linewidth]{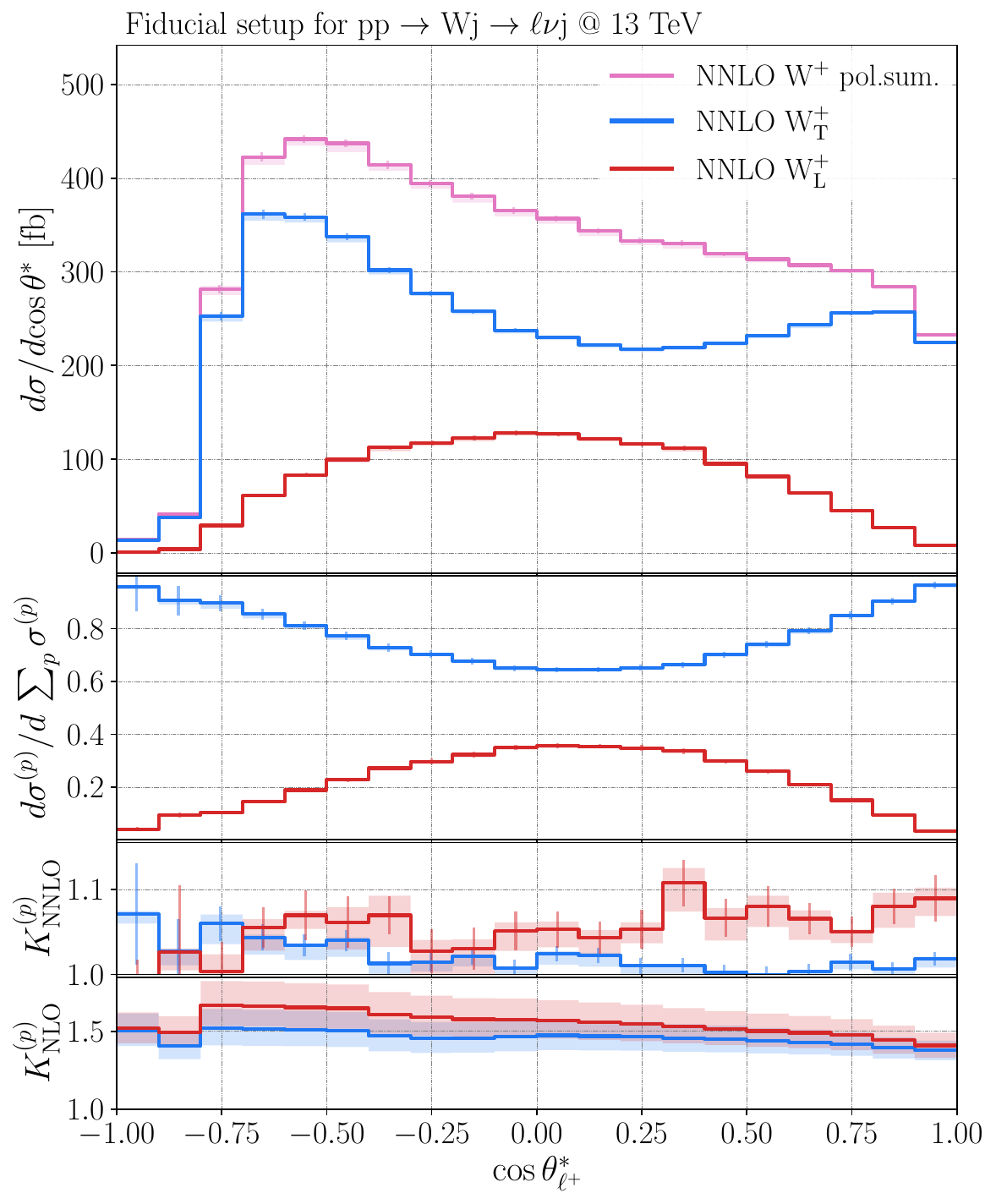}
  \caption{Differential distributions for $\cos\theta^*_{\ell^+}$ in the
    \emph{inclusive} (left) and \emph{fiducial} (right) setup for $\ppmnjP$
    process. In the upper pane, the transverse and longitudinal polarised
    predictions as well as their sum are displayed at NNLO QCD accuracy. In the
    middle pane, the transverse and longitudinal polarised predictions are
    normalised to the sum of the two. The lower pane shows the respective
    $K$-factors (NLO and NNLO) for the various polarised predictions.
    The bands represent the 7-point scale variation while the bars indicate
    the Monte Carlo uncertainty.
    \label{fig:polarisations_cosTheta}
    }
\end{figure}

\interfootnotelinepenalty=10000
We obtain polarisation fractions as a fraction of each polarised cross section
to the overall sum for each setup. Another method to obtain fractions would be
a convolution with Legendre polynomials \cite{Ballestrero:2017bxn} in the
Monte Carlo integration of an unpolarised setup. However,
it is only applicable in the absence of leptonic phase space cuts,
as it relies on the analytic expression of the cross section as a
function of the leptonic emission polar angle\footnote{The
leptonic emission polar angle is defined as the polar angle of
the charged lepton in the $\PW$-boson rest frame, with respect to the
$\PW$-boson flight direction in the laboratory frame.}
\cite{Bern:2011ie,Stirling:2013muo}.
We checked that in the inclusive setup the result of this method coincides
within statistical uncertainties with the last column of \refta{tab:inclusive_xsec}.

\begin{figure}[!t]
  \centering
  \includegraphics[width=.49\linewidth]{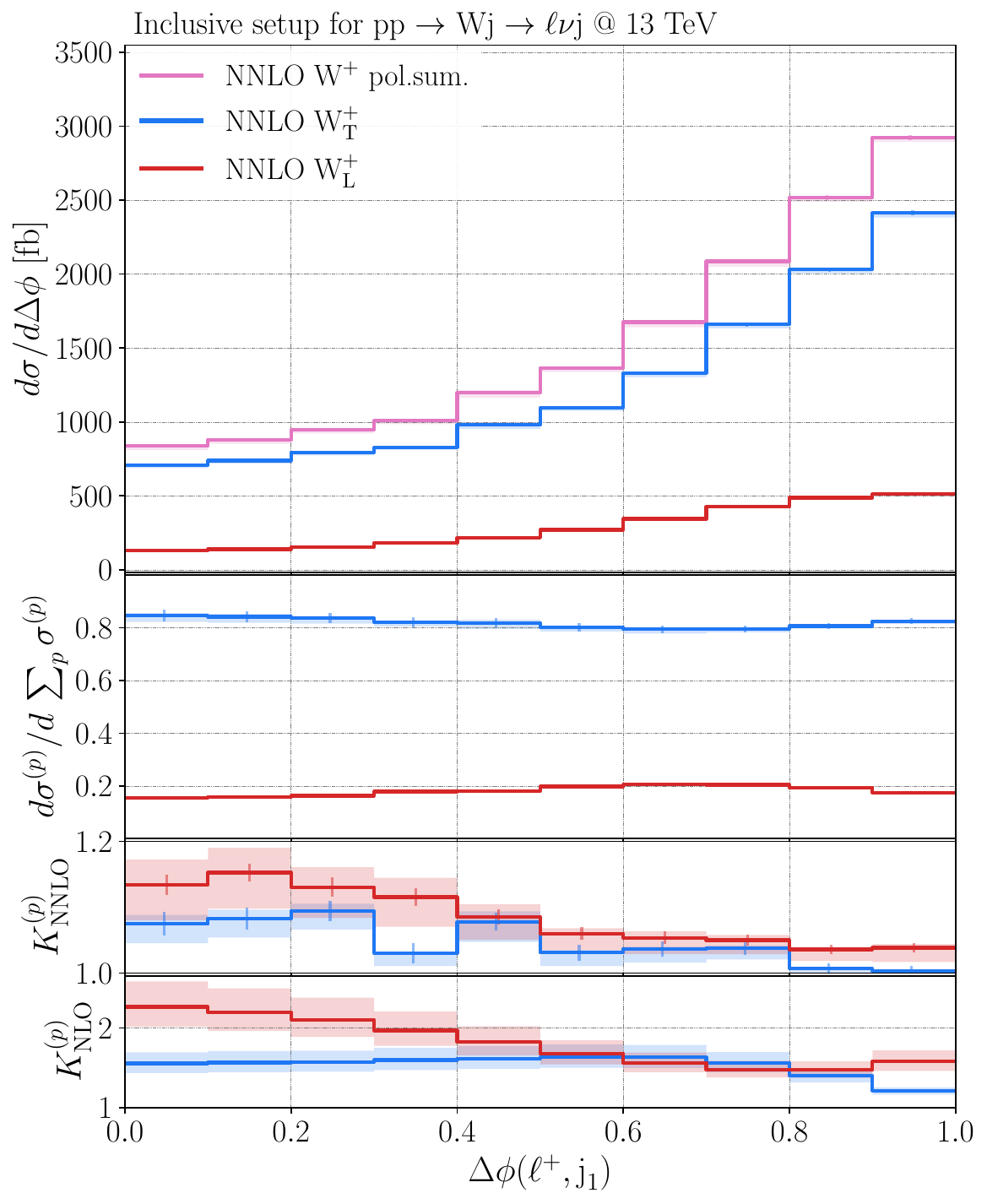}
  \includegraphics[width=.49\linewidth]{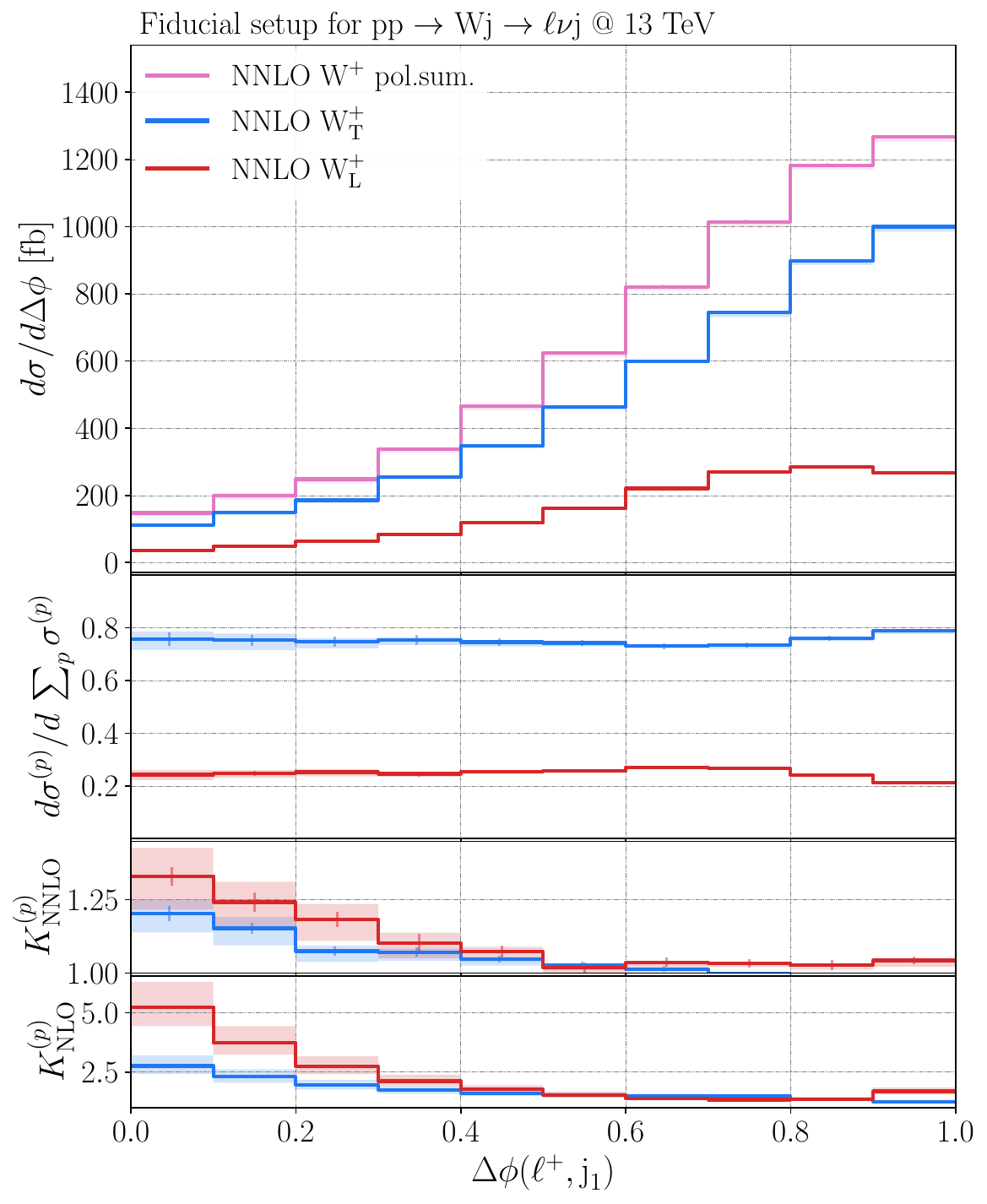}
  \caption{Differential distributions for $\Delta\phi(\ell^+, \Pj_1)$.
    Same plot structure as in \reffi{fig:polarisations_cosTheta}.
    \label{fig:polarisations_philj1}
    }
\end{figure}

Another question is the estimation of theoretical uncertainty for the
polarisation fractions. It was first addressed in \citere{Bern:2011ie}, where
the authors encountered very low sensitivity of the polarisation fractions to the
variation of the scale and resorted to comparing parton-shower and fixed-order
predictions to estimate the uncertainty. We believe this would be the best
strategy for NLO precision. However, with NNLO corrections at hand, we decided
to estimate the scale uncertainty by varying the fixed-order
prediction scale in an uncorrelated way, meaning that errors in the ratio are
propagated via the usual chain rule.

\begin{figure}[!t]
  \centering
  \includegraphics[width=.49\linewidth]{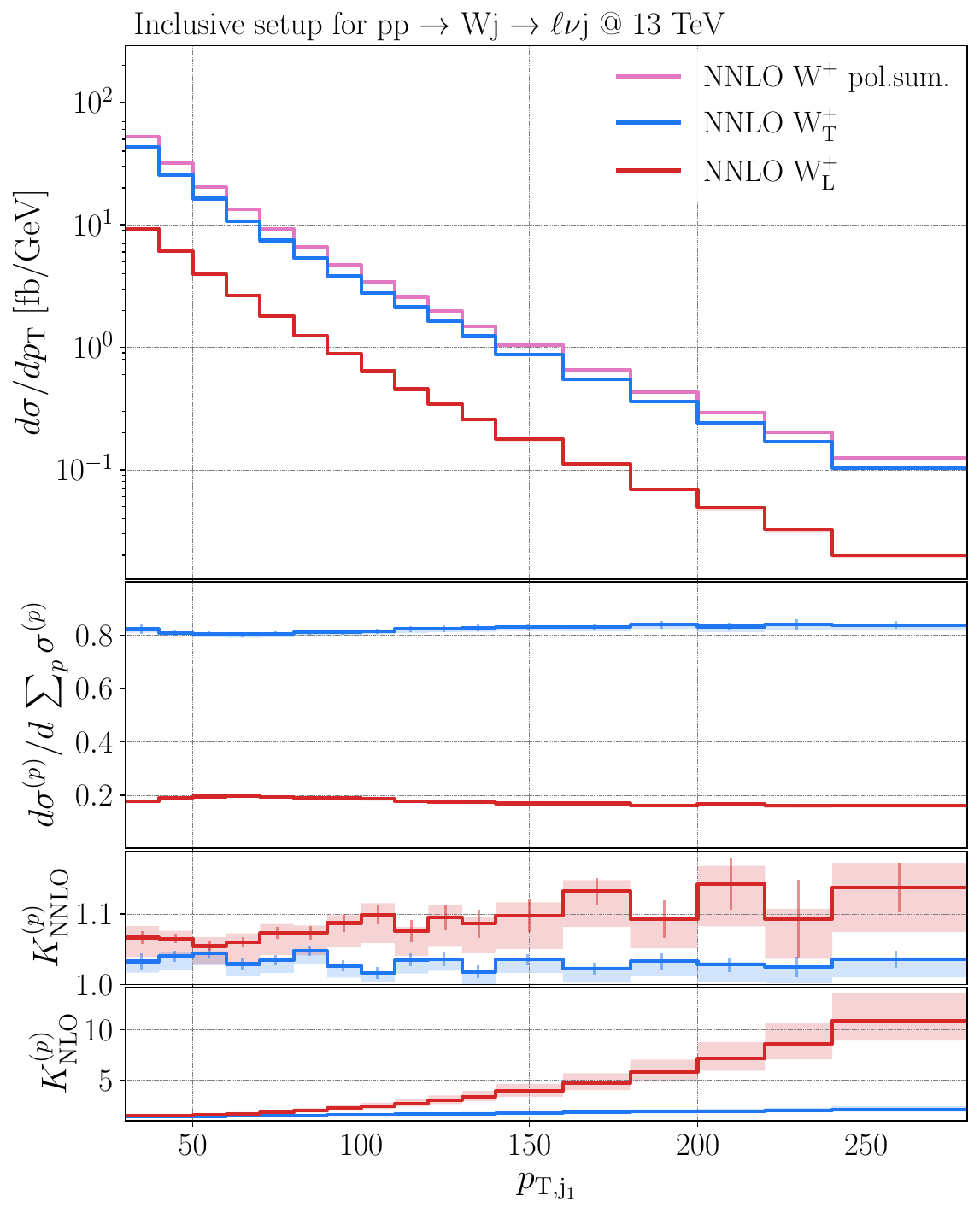}
  \includegraphics[width=.49\linewidth]{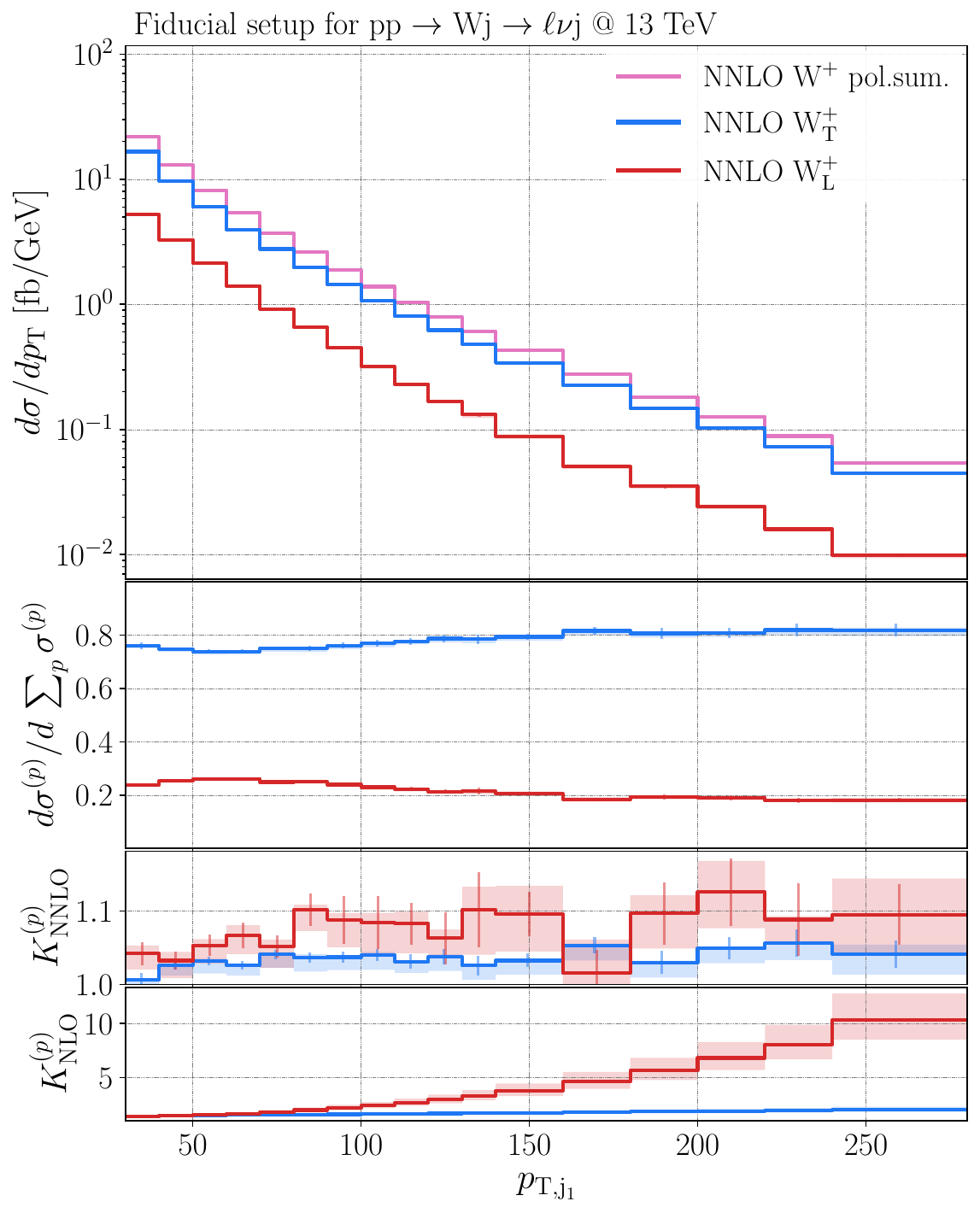}
  \caption{Differential distributions for $p_{\rT, \Pj_1}$.
    Same plot structure as in \reffi{fig:polarisations_cosTheta}.
    \label{fig:polarisations_pTj1}
    }
\end{figure}

\begin{figure}[!t]
  \centering
  \includegraphics[width=.49\linewidth]{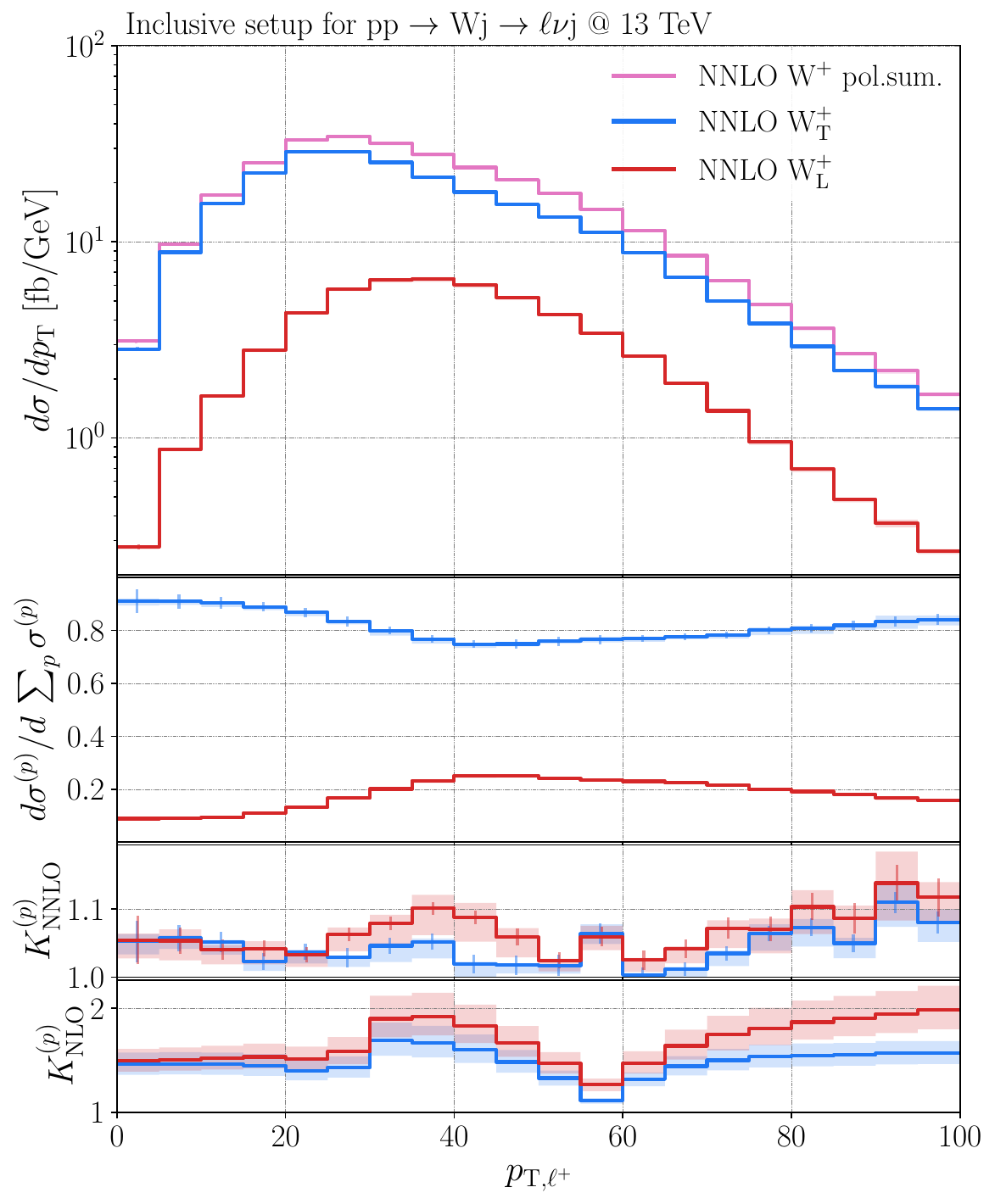}
  \includegraphics[width=.49\linewidth]{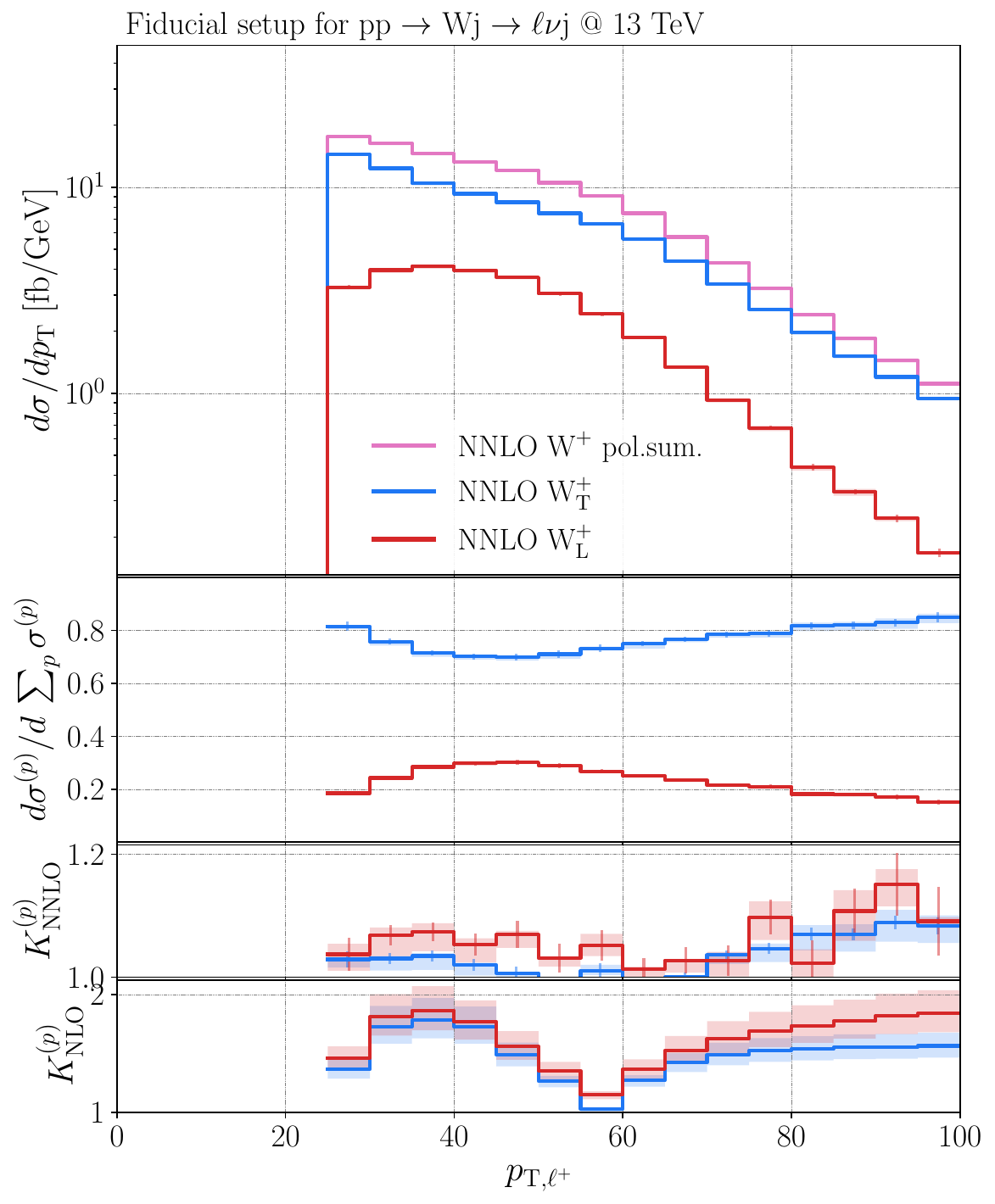}
  \caption{Differential distributions for $p_{\rm T, \ell^+}$.
    Same plot structure as in \reffi{fig:polarisations_cosTheta}.
    \label{fig:polarisations_pTl}
    }
\end{figure}

The effect that clearly stands out is that longitudinal polarisation receives a
larger correction than the transverse one in all setups. It is not
surprising and applies to various other processes
\cite{Poncelet:2021jmj,Denner:2020bcz}. This is even more apparent on several
differential distributions in
figures~\ref{fig:polarisations_cosTheta}-\ref{fig:polarisations_pTl}, where the
plots on the left- and right-hand side represent the inclusive and fiducial
setup, respectively. We only show results for the ``plus'' signature here.

The main observable for vector-boson polarisation is the polar-angle
distribution of the leptonic emission, presented in
\reffi{fig:polarisations_cosTheta}. In the inclusive setup the longitudinal
mode is perfectly symmetric in the angle of emission; the transverse mode
dominates and peaks for the case of backwards emission, as a consequence of
W-bosons being created left-handed at the LHC \cite{Bern:2011ie}.
Fiducial cuts significantly
change the shape of the absolute distributions. However, this does not have a
large effect on the distributions of the fractions. Due to their distinct
profiles, these shapes would allow
for the most efficient separation of polarisations via shape fitting, but alas,
the neutrino momentum is not available experimentally to fully reconstruct the
angle of leptonic emission.  Therefore, experiments have utilised available
transverse momenta to have a handle on the emission angle.  These are
$\cos\theta^*_{\text{2D}}$ for ATLAS \cite{ATLAS:2012au} and $L_P$ for CMS
\cite{CMS:2011kaj}.  Naturally, the difference in shapes between polarisations
is partially washed out in comparison with the actual polar-angle distribution.

An angular observable that is readily available at the experimental level is for
example, the azimuthal angle between the charged lepton and the hardest jet ($\Pj_1$),
presented in \reffi{fig:polarisations_philj1}. This observable was measured,
for example, by the CMS experiment in ref.~\cite{CMS:2017gbl}.
The corrections at NLO are different
between the longitudinal and transverse polarisations but are quite similar at NNLO.
Fiducial cuts significantly raise the QCD corrections at lower angles.

QCD corrections are particularly important in energy related distributions.
In \reffi{fig:polarisations_pTj1} we display transverse momentum
of the hardest jet. The NLO corrections are rising seemingly without a bound
for the longitudinal component.
At NNLO, the corrections are stable,
oscillating around 10\%, showing a good sign for the perturbative convergence.
Similar giant $K$-factor at NLO appear in other energy-related
distributions involving jets: $H_\rT$, invariant masses,
other $p_\rT$ distributions.

In the context of shape fits, we would like to discuss the observable
for the transverse momentum of the charged lepton in
\reffi{fig:polarisations_pTl}, although similar features
are naturally present in the missing $p_\rT$ distribution. It has a remarkably
distinct shape of polarisation fraction profile, which is useful for shape
fitting. The NLO corrections are typically large at high $p_\rT$ but seem
to flatten out in contrast to the case of jet $p_\rT$ highlighted above.
The shapes of
corrections seem similar otherwise. The longitudinal polarisation receives
stronger corrections near the peak, reflecting the fact that at the level of
total cross section the $K$-factors are larger.

\begin{figure}[!b]
  \centering
  \includegraphics[width=.99\linewidth]{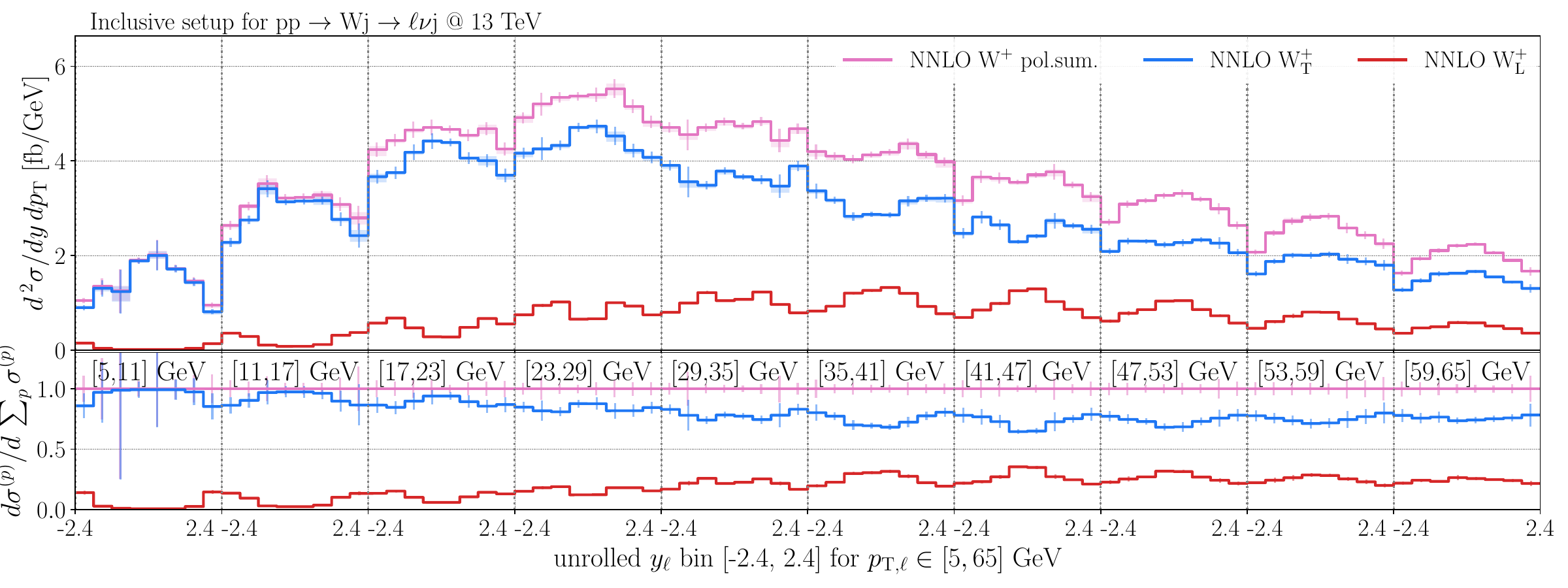}

  \caption{
    Double differential distributions for $\ppmnjP$ in the \emph{inclusive}
    setup in the $p_{\rm T, \ell^+}$ and $y_{\ell^+}$ observables. In the upper
    plot, the transverse and longitudinal polarised predictions as well as
    their sum are displayed at NNLO QCD accuracy. In the lower pane, the
    transverse and longitudinal polarised predictions are normalised to the sum
    of the two. The bands represent the 7-point scale variation
    while the bars indicate the Monte Carlo uncertainty.
      \label{fig:2D}
    }
\end{figure}

\begin{figure}[!t]
  \centering
  \includegraphics[width=.49\linewidth]{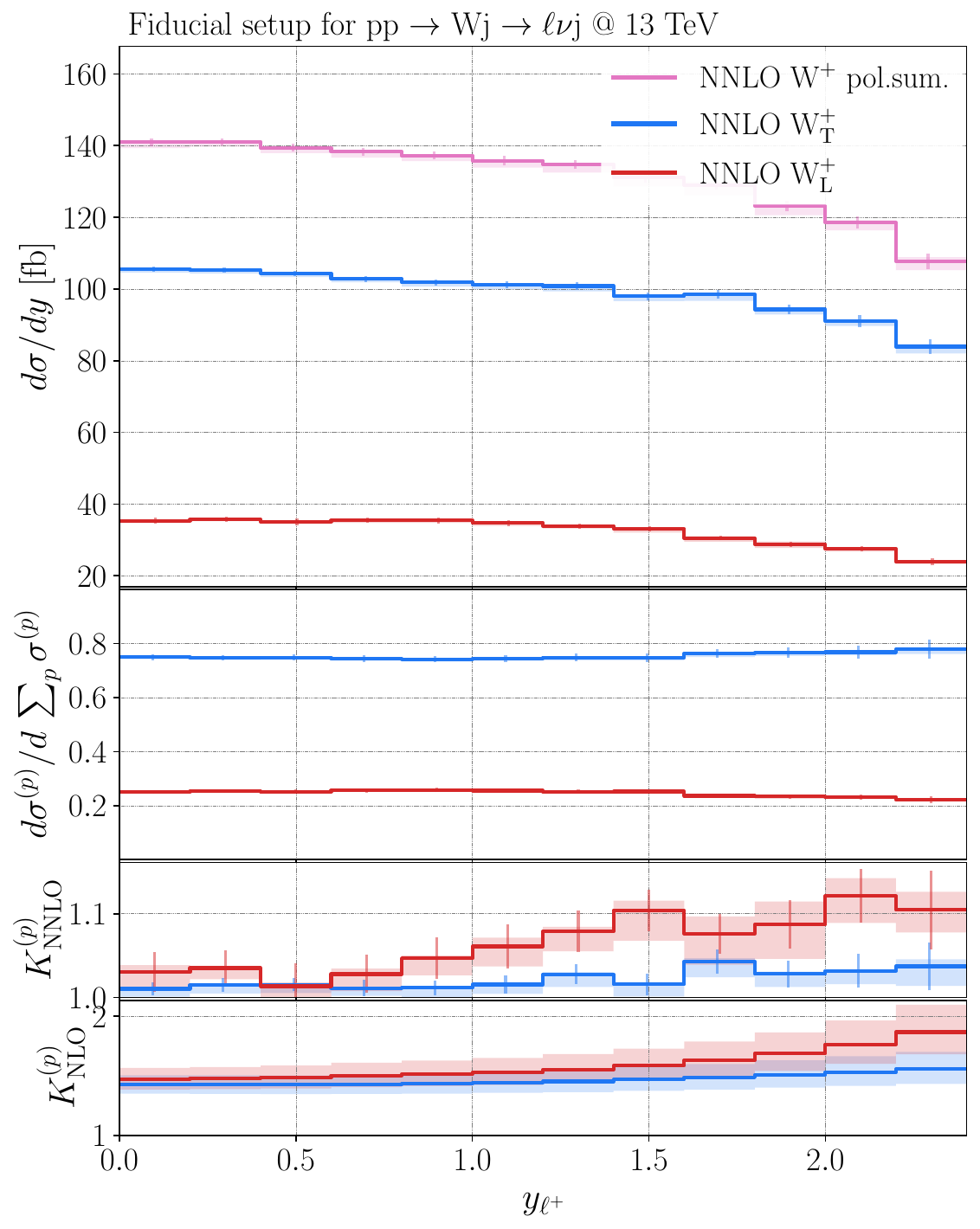}
  \includegraphics[width=.49\linewidth]{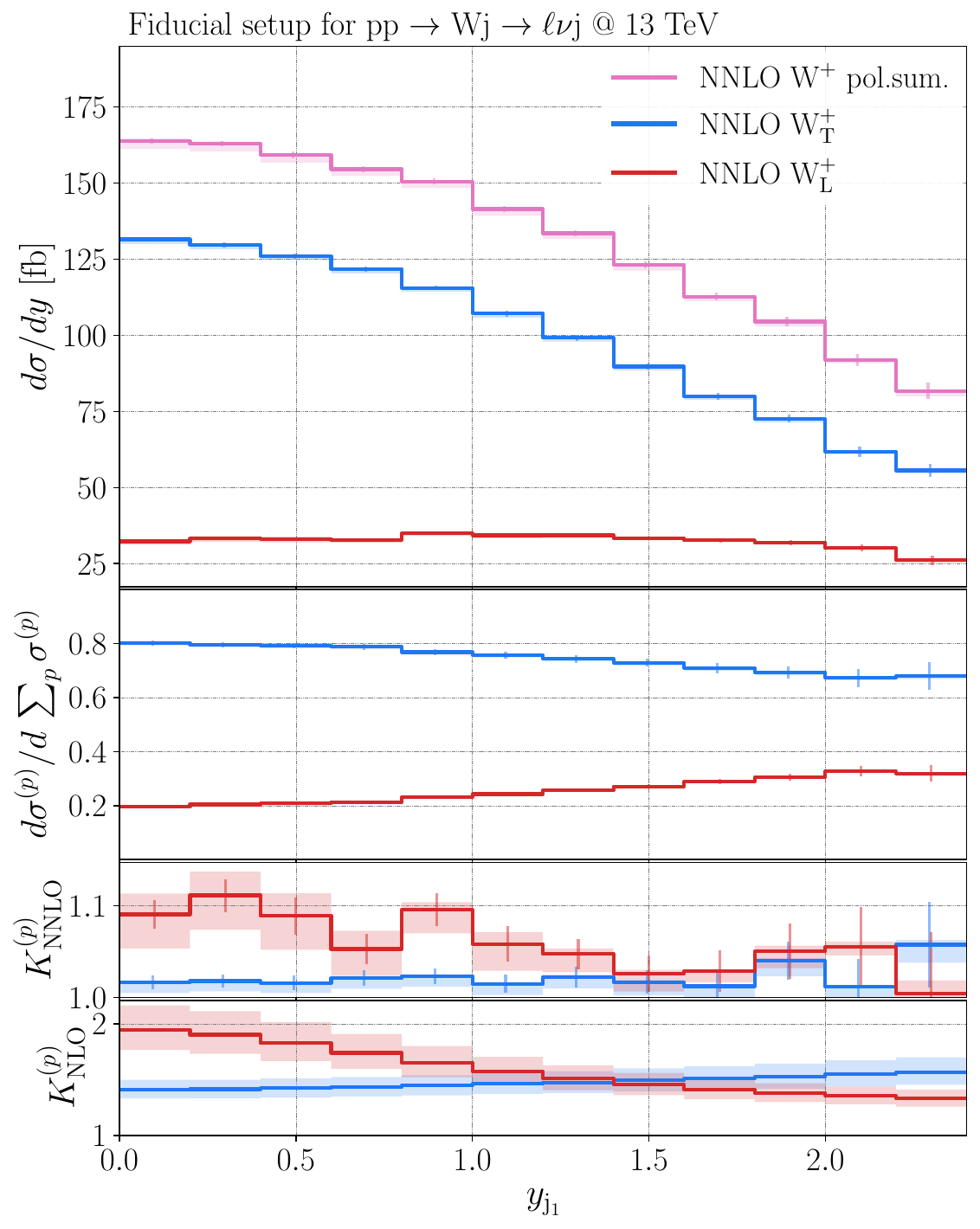}
  \caption{Differential distributions for symmetrised $y_{\ell^+}$ (left) and $y_{\Pj_1}$ (right), both in the \emph{fiducial} setup.
    Individual plots follow the structure of \reffi{fig:polarisations_cosTheta}.
    \label{fig:polarisations_yj1}
    }
\end{figure}

\begin{figure}[!t]
  \centering
  \includegraphics[width=.49\linewidth]{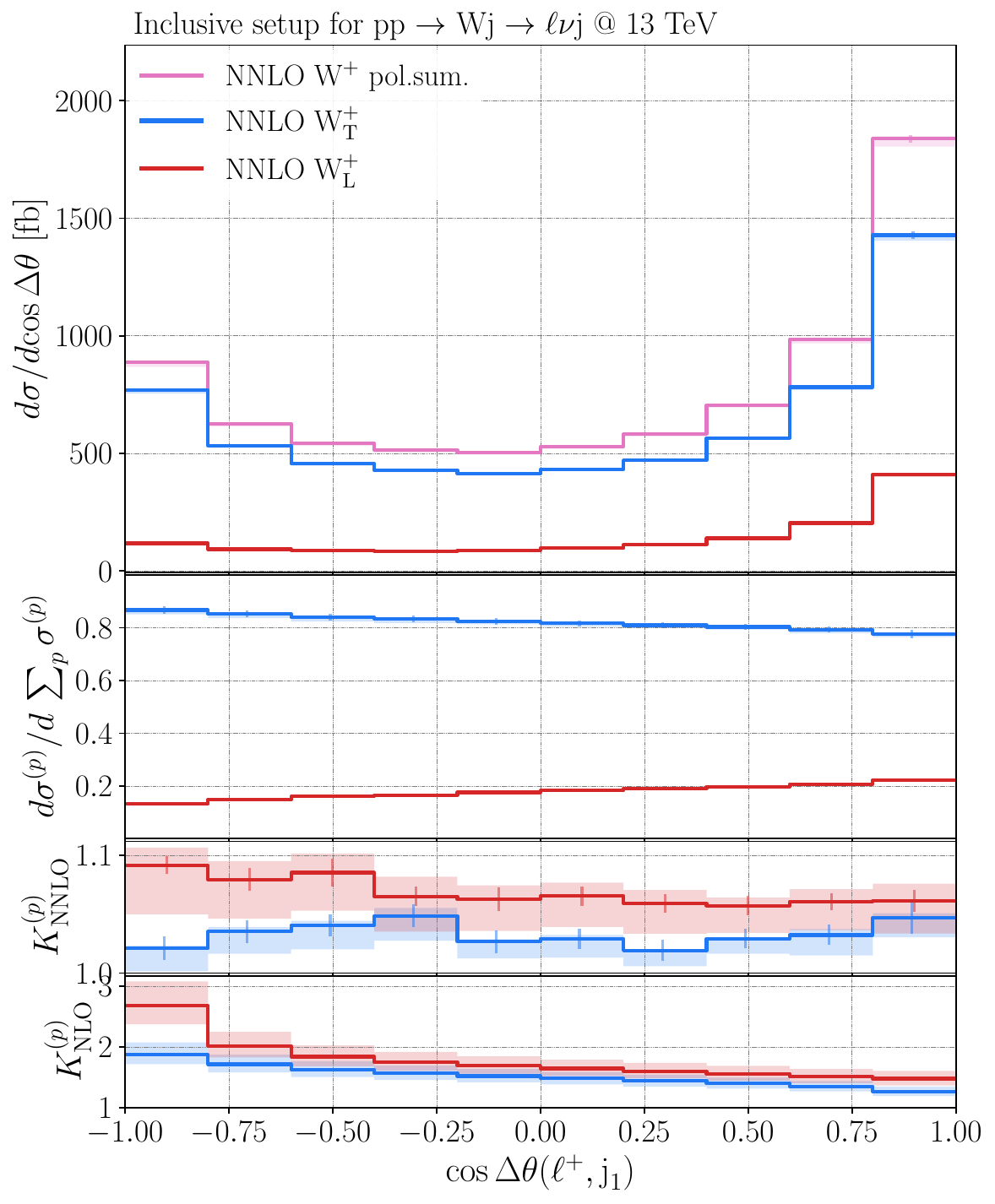}
  \includegraphics[width=.49\linewidth]{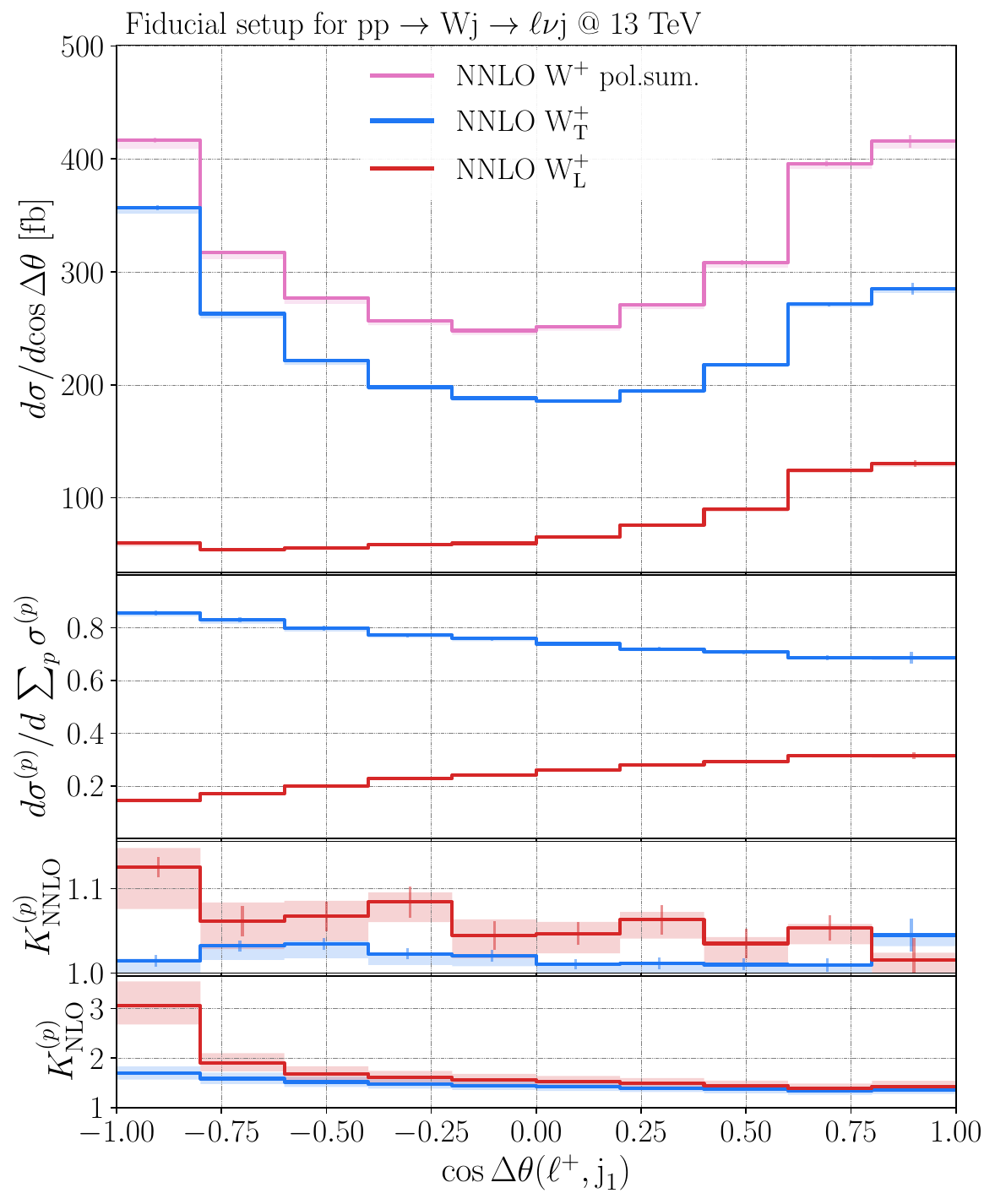}
  \caption{Differential distributions for symmetrised $\cos\theta(\ell^+,\Pj_1)$
    in the \emph{inclusive} (left) and \emph{fiducial} (right) setups.
    Individual plots follow the structure of \reffi{fig:polarisations_cosTheta}.
    \label{fig:polarisations_cosTlj1}
    }
\end{figure}

In \reffi{fig:2D}, we show the double
differential distribution in the transverse momentum and the rapidity of the
charged lepton. Such a 2D information can also be particularly useful for the
extraction of polarisation fractions through fitting as they provide complementary
information. This kind of measurements have been already presented in
ref.~\cite{CMS:2020cph} for the Drell-Yan process. In the present case, the transverse momentum
and the rapidity of the charged lepton constitute a very good combination
as they are not correlated.
The 2D distribution features a change
of rapidity profiles for different transverse momentum bins.
Interestingly, the ordinary 1D lepton rapidity distribution shows
almost completely flat polarisation fraction profile.
One can observe that the longitudinal polarisation is at maximum for transverse
momentum around $50\GeV$ and central rapidity. An experimental $\Wplusj$
analysis based on 2D information would be therefore highly valuable and very
complementary to the one for the Drell-Yan process.

A possibly even more sensitive observable in combination with lepton $p_\rT$
would make the jet rapidity, which
we compare separately to lepton rapidity in \reffi{fig:polarisations_yj1}.
Jet rapidity seems to have more pronounced monotonically decreasing
polarisations shapes at lower rapidities $\abs{y}<1.7$ ---
which is the area where off-shell and interference
effects are low, as is shown further in section \ref{sec:OS_IE_data}.
Similarly to lepton rapidity, jet rapidity is (to a large extent)
independent of lepton $p_\rT$, and thus
if one wants to make a more constraining fit, the possible alternative
would be to use a 2D distribution of jet rapidity and lepton $p_\rT$,
which will one could in principle fit all polarisations and signatures
independently at once.

Finally, we show another observable in \reffi{fig:polarisations_cosTlj1}
which we believe would be
helpful in experimental analyses: the angle between the charged lepton
and the hardest jet. The difference between polarisation shapes is by a factor
of 2 more pronounced in the fiducial setup, which makes it very useful
in extracting polarisations using the method of shape fitting.
The NNLO corrections for both polarisations are mild and flat.

\subsection{Charge differences}
\label{sec:charges}

The two signatures of the $\ppmnj$ process behave differently,
and there are two distinctive reasons within the SM for that:
differences in the lepton emission distribution, and the PDFs
of the initial-state partons.

\begin{figure}[!t]
  \centering
  \includegraphics[width=.49\linewidth]{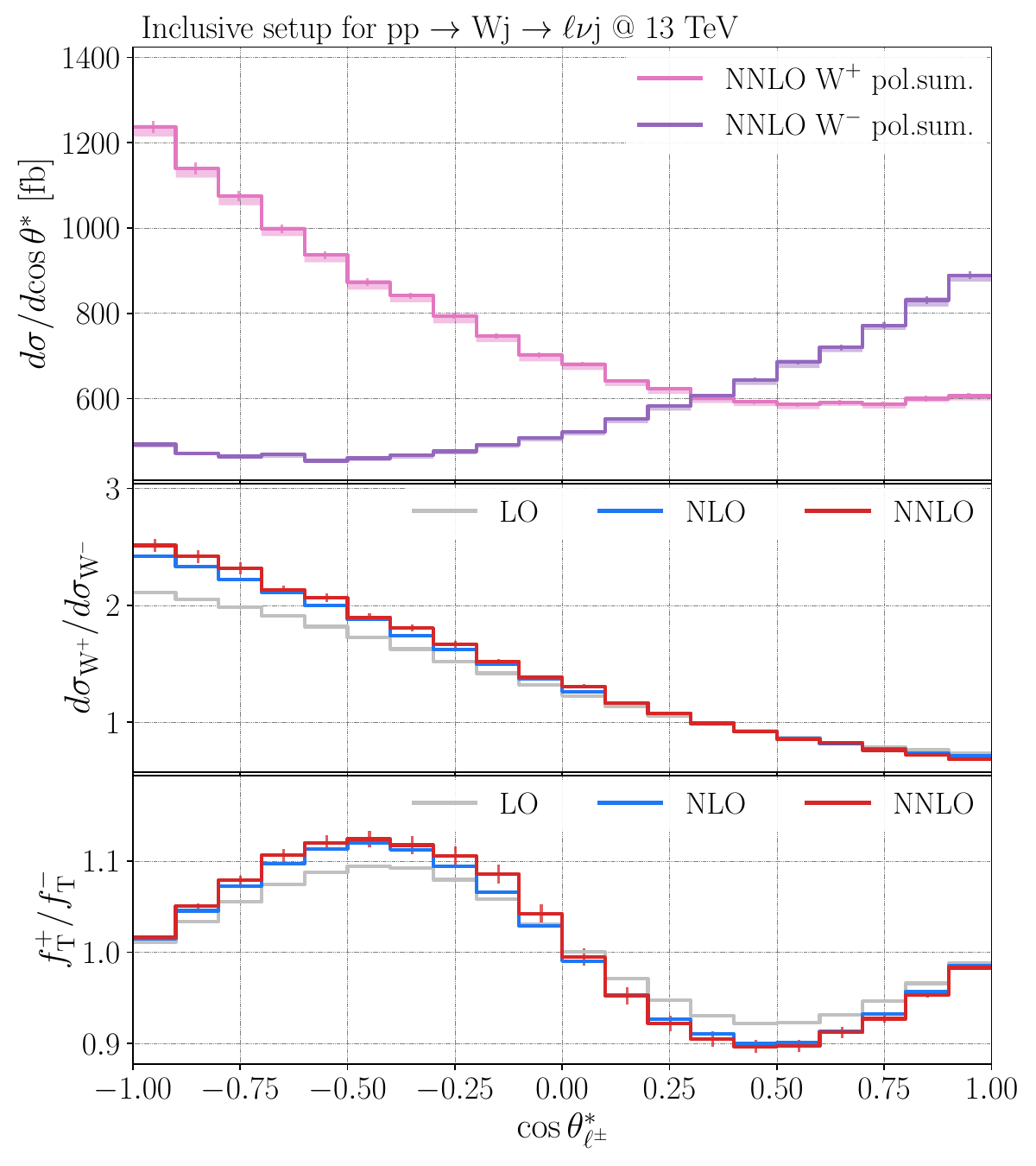}
  \includegraphics[width=.49\linewidth]{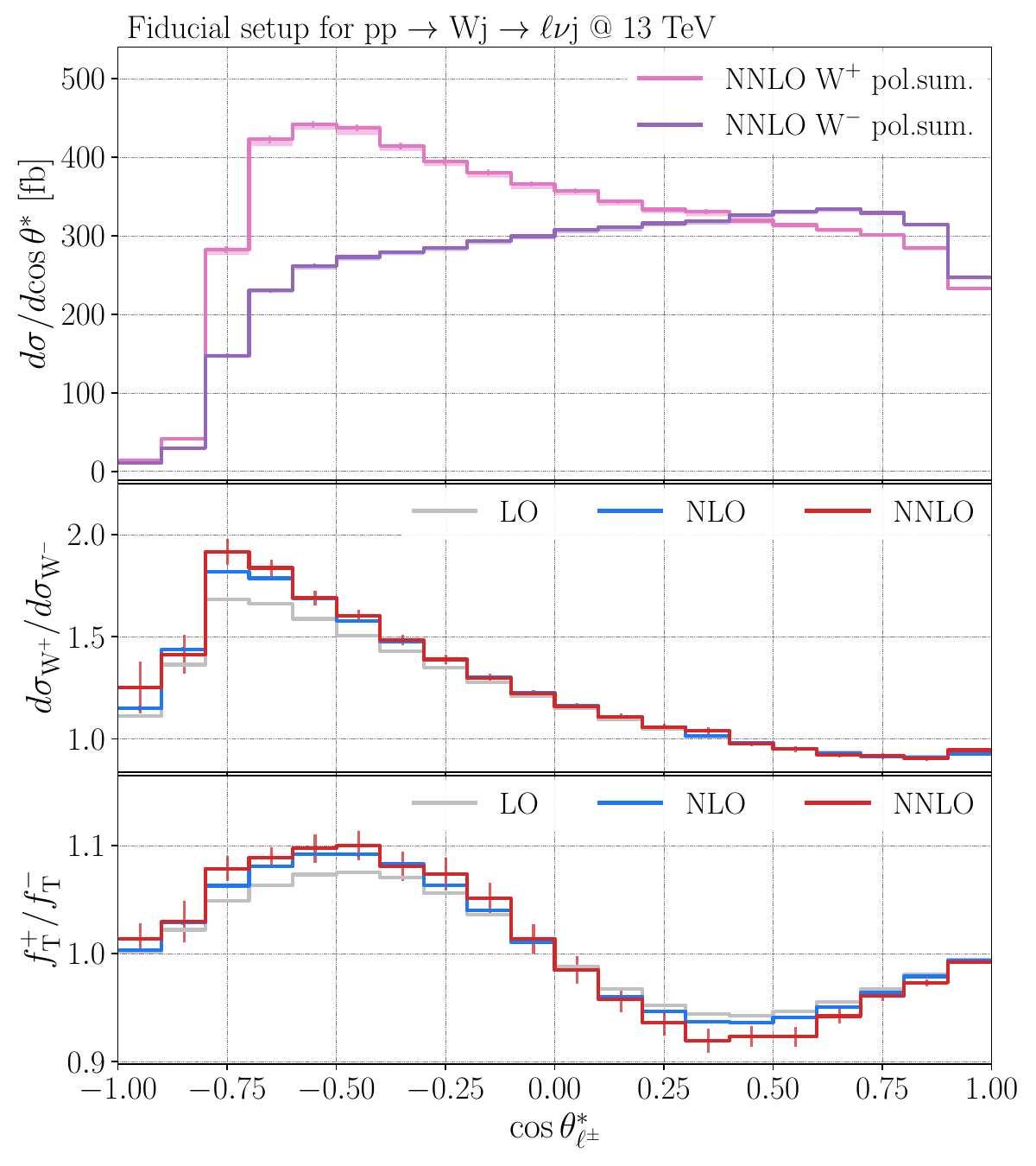}
  \caption{Differential distributions of $\cos \theta^*_{\ell^\pm}$
    for $\ppmnj$
    in the \emph{inclusive} (left) and \emph{fiducial} (right) setup.
    In the upper pane, the sum of the transversely and longitudinally polarised
    predictions for both signatures are displayed at NNLO QCD accuracy. In the
    middle pane, the ratio of the two signatures is shown at LO, NLO, and
    NNLO.
    The lower pane presents the ratio of the differential distributions of
    transverse polarisation fractions
    between the two signatures, also at LO, NLO, and NNLO. In the upper plot, the bands
    represent the 7-point scale variation, while the bars indicate the Monte
    Carlo uncertainty for in all three plots.
      \label{fig:charges_cosTheta}}
\end{figure}

\begin{figure}[!t]
  \centering
\includegraphics[width=.49\linewidth]{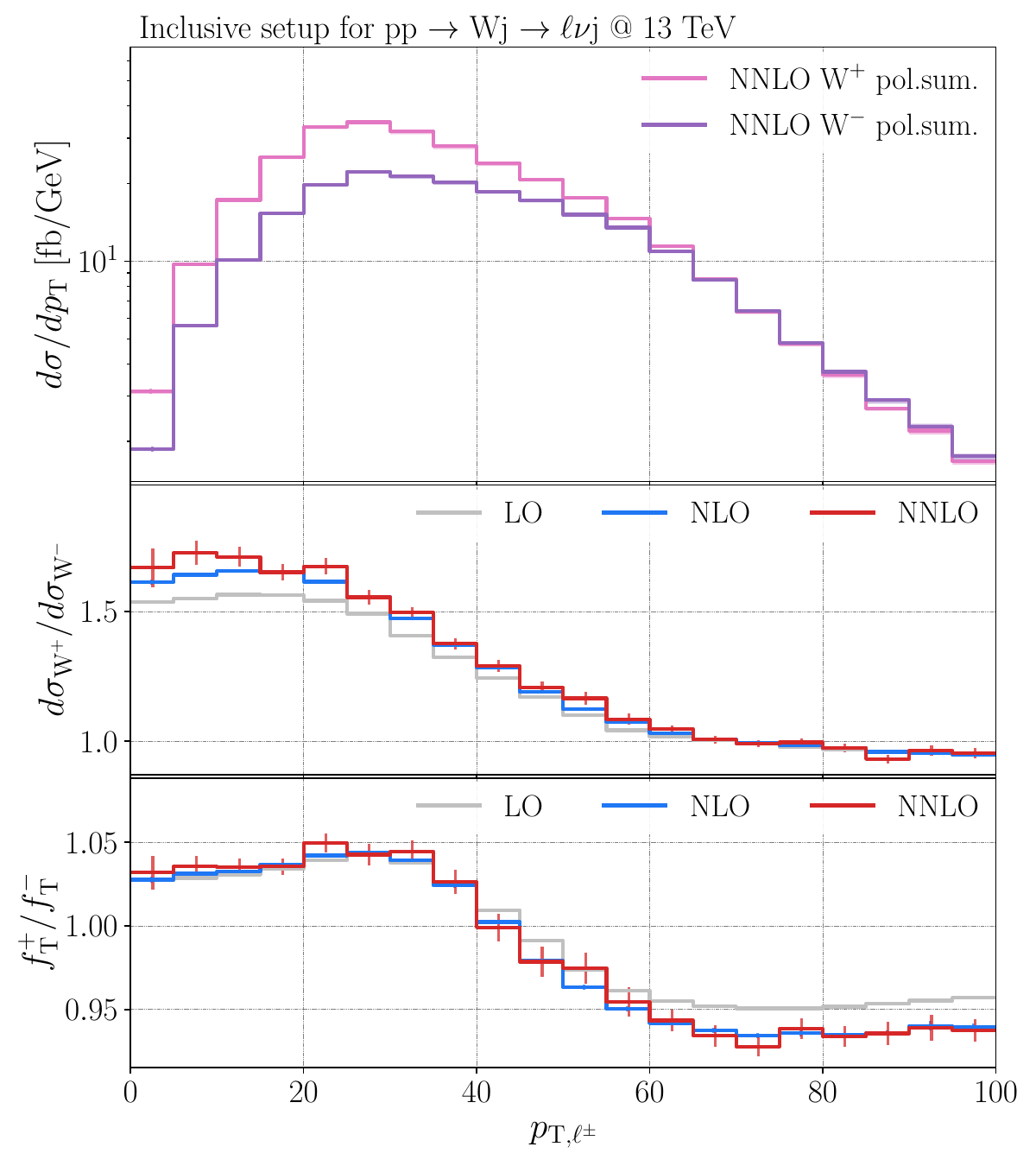}
\includegraphics[width=.49\linewidth]{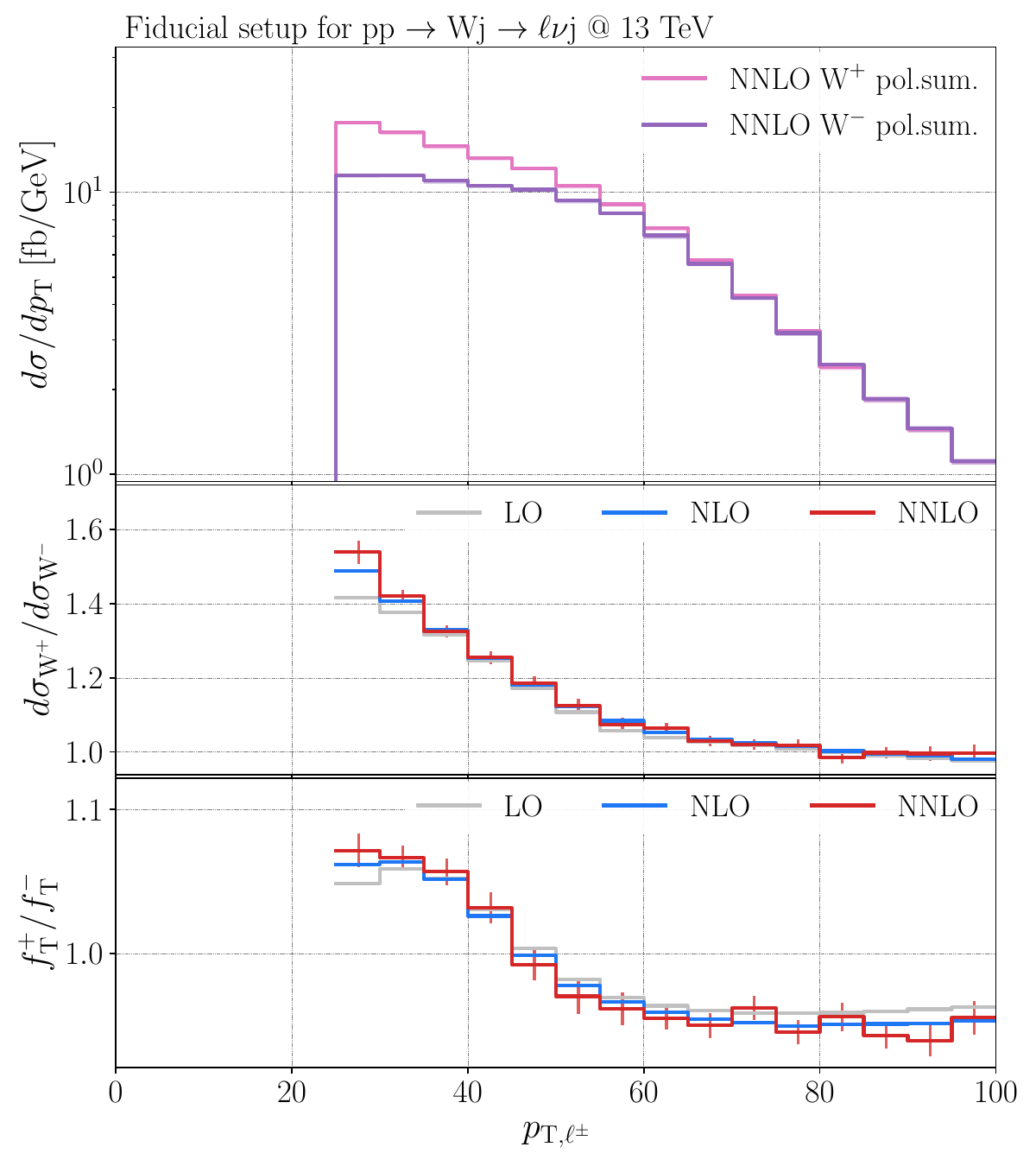}
    \caption{Differential distributions of $p_{\rT}(\ell)$ for $\ppmnj$
      in the \emph{inclusive} (left) and \emph{fiducial}
      (right) setup.  same individual plot structure as in
      \reffi{fig:charges_cosTheta}.
      \label{fig:charges_pTl}}

\vspace{1em}
\includegraphics[width=.49\linewidth]{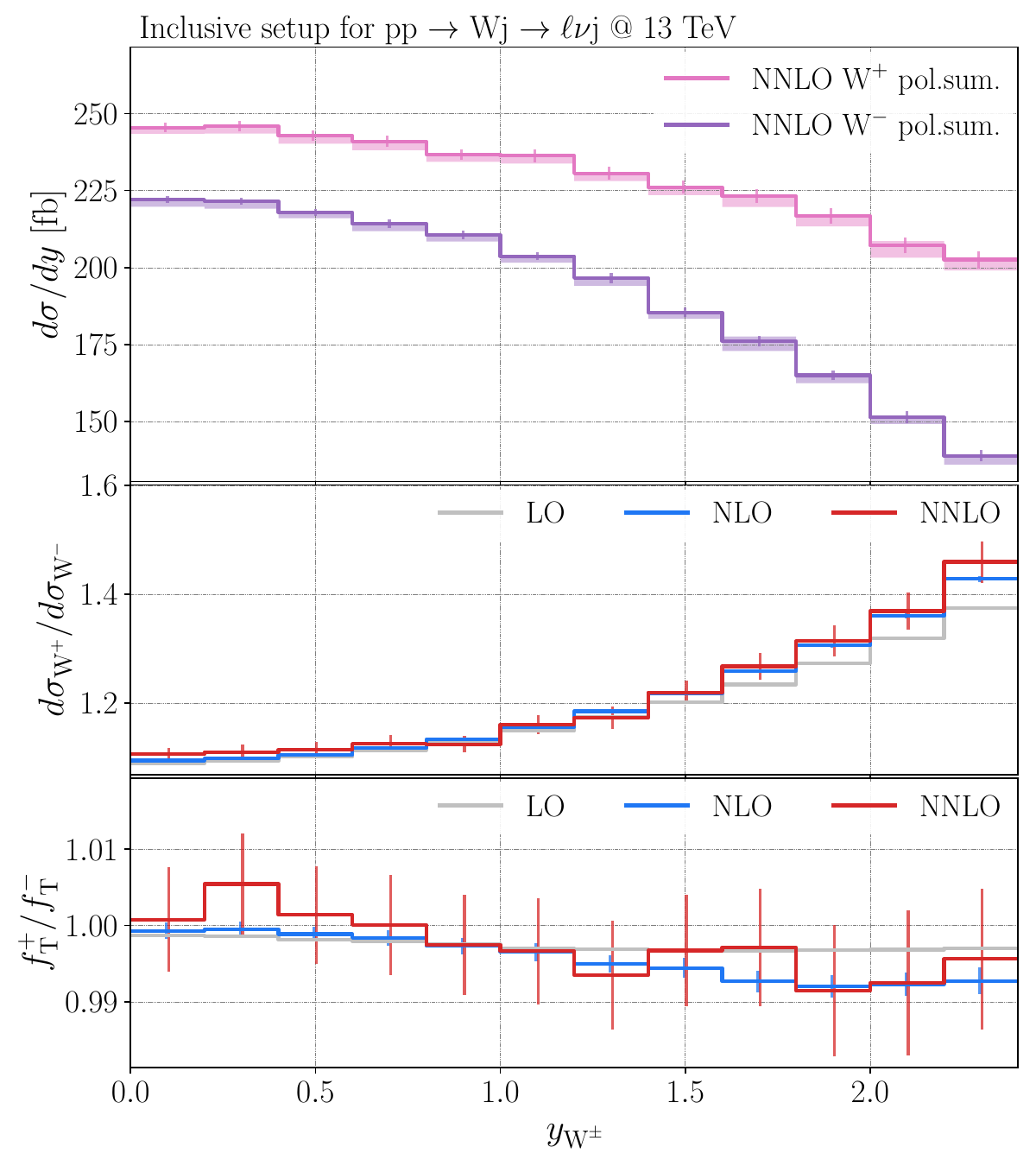}
\includegraphics[width=.49\linewidth]{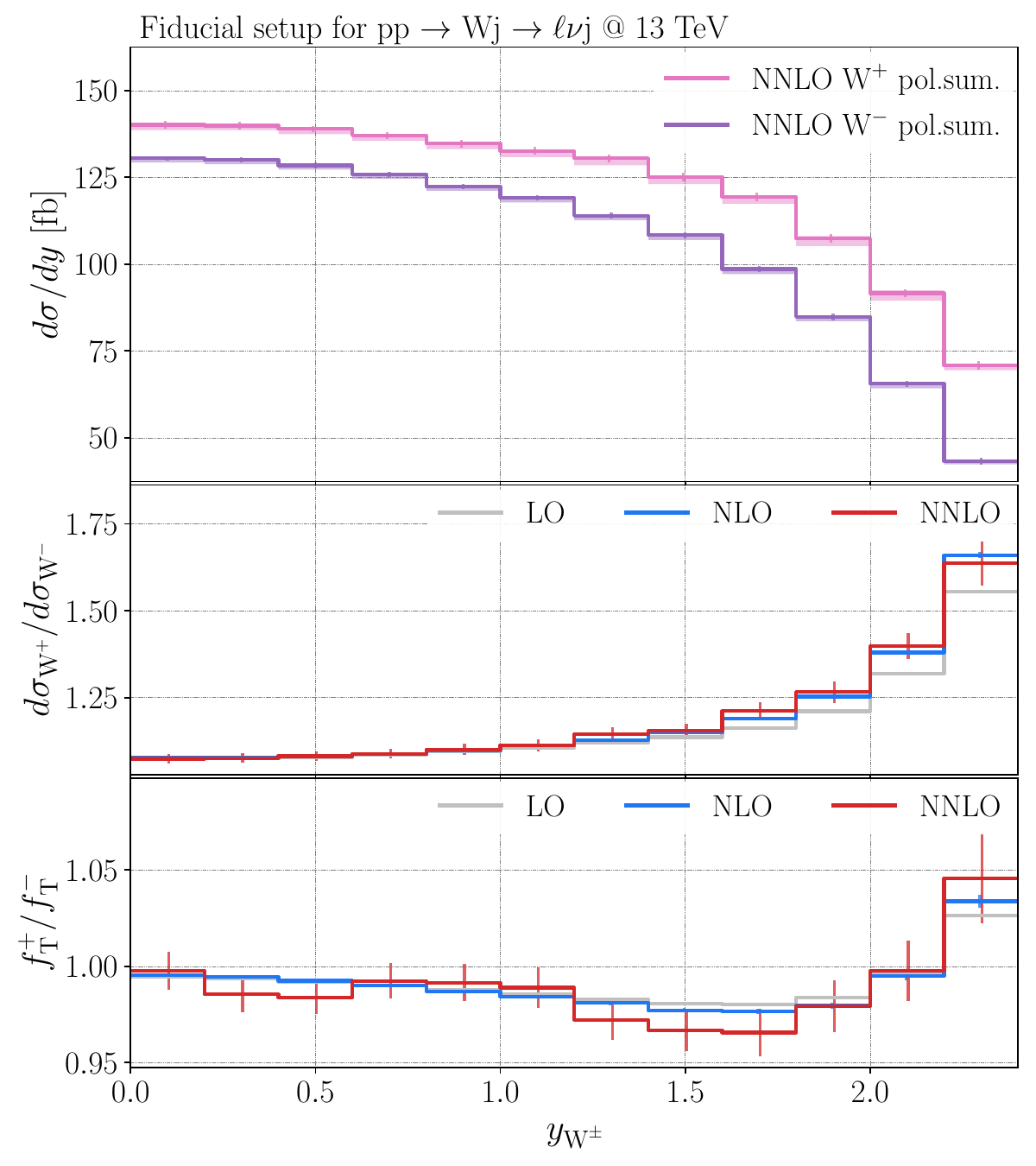}
  \caption{Differential distributions of $y_{\PW}$ for $\ppmnj$
    in the \emph{inclusive} (left) and \emph{fiducial} (right) setup.
    Same individual plot structure as in \reffi{fig:charges_cosTheta}.
      \label{fig:charges_yW}}
\end{figure}

The first effect is evident from the distribution of the polar emission
angle of the charged lepton in \reffi{fig:charges_cosTheta}.
In the inclusive setup, the shapes look almost mirrored,
and peak in opposite the regions:
at $-1$ for the ``plus'' signature, and at $+1$ for the ``minus'' one.
In fact, these shapes can be described analytically, as long as
no leptonic cuts are imposed \cite{Bern:2011ie}.
The shapes change dramatically on the edges in the fiducial setup
due the kinematic constrains on the lepton.
It is interesting to notice that the ratio of the polarisation fraction
of the two signatures shows a perfect oscillatory behaviour that can
be traced back to their analytical expressions and is thus independent of the setup.
Also, it is worth mentioning that such ratios
(see also \reffis{fig:charges_pTl} and \ref{fig:charges_yW})
show essentially no shape distortion with respect to higher-order corrections,
making them particularly attractive for experimental studies.

The direction of emission angle is directly connected to the transverse
momentum of the lepton, displayed in \reffi{fig:charges_pTl}. We observe a
constant behaviour except in the region $25 <
p_{\rT,\ell} < 60 \GeV$, where ``minus'' signature has a much flatter behaviour
and catches up with ``plus'' signature at the beginning of the tail. As the
emission angle distribution shows, the lepton of the ``minus'' signature tends
to be emitted in the direction of the W-boson, thus making it more likely to
have higher $p_\rT$. Fiducial cuts do not affect these conclusions, except that
there are no events below the leptonic cut. This observable also has an
excellent capability to discern not only polarisations but also the particular
process signature.

The other effect is due to differences in PDFs for the initial-state partons. As
was noted previously, the ``plus'' signature has a larger cross-section due to
the dominance of ug-channel over dg-channel. The distribution of W-boson
rapidity in \reffi{fig:charges_yW} also looks much more spread out for the
``plus'' signature. The same effect is present, although in a more suppressed
form, in $y_\ell$ and $y_{\text{j}_1}$. As for the ratio of transverse
polarisation fractions, one can see that this observable is not sensitive to
left-right asymmetry of W-bosons.

\subsection{Off-shell, interference effects, and comparison with data}
\label{sec:OS_IE_data}

So far, the discussion focused on polarised predictions. Nonetheless, such
predictions are only approximations to the full off-shell and unpolarised
process. In this section we address limitations of the NWA polarised computation by
comparing it to unpolarised predictions (within the NWA framework)
as well as to the fully off-shell calculation.
We also compare simulations with the available data from the CMS
collaboration \cite{CMS:2017gbl}.

We start by comparing unpolarised predictions to the sum of the polarised
ones in the inclusive setup in \refta{tab:inclusive_xsec_approx}.
The difference between the two constitutes an estimate of the interference effects
between the polarisation states. Interference effects appear to be
consistent with zero within the Monte Carlo uncertainty for all perturbative
orders and for both signatures.

\begin{table}[!t]
  \centering
  {\small
  \renewcommand{\arraystretch}{1.3}
  \begin{tabular}{|C{2.5cm}|C{2.26cm}C{2.26cm}C{0.95cm}C{2.16cm}C{0.95cm}|}
    \hline
        \cellcolor{blue!9}{\textit{Inclusive}}
      & \cellcolor{blue!9}{LO [fb]}
      & \cellcolor{blue!9}{NLO [fb]}
      & \cellcolor{blue!9}{$K_\text{NLO}$}
      & \cellcolor{blue!9}{NNLO [fb]}
      & \cellcolor{blue!9}{$K_\text{NNLO}$}
      \\
    \hline %
    $\PW^+_{\text{unpol.\ }}$
        & $990.00(8)^{+11.5\%}_{-9.3\%}$
        & $1483.1(5)^{+7.9\%}_{-7.0\%}$
        & $1.50$
        & $1541(5)^{+0.5\%}_{-1.9\%}$
        & $1.04$
     \\
    $\PW^+$ pol.\ sum.\
        & $989.97(7)^{+11.5\%}_{-9.3\%}$
        & $1482.8(5)^{+7.9\%}_{-7.0\%}$
        & $1.50$
        & $1546(6)^{+0.3\%}_{-1.9\%}$
        & $1.04$
     \\
    \hline %
    $\PW^-_{\text{unpol.\ }}$
        & $774.04(6)^{+11.5\%}_{-9.3\%}$
        & $1121.1(3)^{+7.1\%}_{-6.5\%}$
        & $1.45$
        & $1158(4)^{+0.3\%}_{-1.7\%}$
        & $1.03$
     \\
    $\PW^-$ pol.\ sum.\
        & $773.97(5)^{+11.5\%}_{-9.3\%}$
        & $1121.2(3)^{+7.1\%}_{-6.5\%}$
        & $1.45$
        & $1155(3)^{+0.2\%}_{-1.5\%}$
        & $1.03$
     \\
     \hline
  \end{tabular}
  }
  \caption{Total cross sections of $\ppmnj$ for the \emph{inclusive} setup at
    the LHC with the centre-of-mass energy of $\sqrt{s}=13\TeV$. The unpolarised and
    polarised-sum predictions are provided at LO, NLO, and NNLO accuracy
    along with the corresponding $K$-factors.
    Numbers in parentheses indicate statistical errors
    while sub- and superscripts represent scale uncertainty.}
  \label{tab:inclusive_xsec_approx}
\end{table}

Similarly, we provide predictions for the fiducial setup in
\refta{tab:fiducial_xsec_approx}, where we also compare to off-shell results.
The difference between the unpolarised and the sum of polarised results
indicates negligible (per mille) interference effects. The difference to
off-shell computation however is much more significant and presents a 1-2\%
effect. This is in agreement with the naive estimate
$\ordercoupling{\Gamma_\PW/M_\PW}$ for the accuracy of the NWA. We note that these
differences are purely due to off-shell effects and do not involve any missing
non-resonant diagrams in the NWA calculation, which is the case for other
processes involving neutral currents. Off-shell effects persist throughout QCD
corrections as they are only concerned with the production part of the amplitude,
whereas the decay part is purely an EW process.

\begin{table}[!h]
  \centering
  {\small
  \renewcommand{\arraystretch}{1.3}
  \begin{tabular}{|C{2.5cm}|C{2.26cm}C{2.26cm}C{1.15cm}C{2.26cm}C{1.15cm}|}
    \hline
        \cellcolor{blue!9}{\textit{Fiducial}}
      & \cellcolor{blue!9}{LO [fb]}
      & \cellcolor{blue!9}{NLO [fb]}
      & \cellcolor{blue!9}{$K_\text{NLO}$}
      & \cellcolor{blue!9}{NNLO [fb]}
      & \cellcolor{blue!9}{$K_\text{NNLO}$}
      \\
    \hline
    $\PW^+$ off-shell
        & $408.69(3)^{+11.5\%}_{-9.3\%}$
        & $607.7(4)^{+7.1\%}_{-6.5\%}$
        & $1.49$
        & $626(3)^{+0.2\%}_{-1.6\%}$
        & $1.03$
     \\
    $\PW^+_{\text{unpol.\ }}$
        & $413.83(2)^{+11.5\%}_{-9.3\%}$
        & $615.7(2)^{+7.1\%}_{-6.5\%}$
        & $1.49$
        & $635(2)^{+0.2\%}_{-1.6\%}$
        & $1.03$
     \\
    $\PW^+$ pol.\ sum.\
        & $413.21(3)^{+11.5\%}_{-9.3\%}$
        & $613.6(2)^{+7.1\%}_{-6.5\%}$
        & $1.49$
        & $633(2)^{+0.2\%}_{-1.6\%}$
        & $1.03$
     \\
    \hline
    $\PW^-$ off-shell
        & $347.19(2)^{+11.5\%}_{-9.3\%}$
        & $504.8(1)^{+6.6\%}_{-6.1\%}$
        & $1.45$
        & $518(2)^{+0.3\%}_{-1.3\%}$
        & $1.03$
     \\
    $\PW^-_{\text{unpol.\ }}$
        & $352.94(2)^{+11.5\%}_{-9.3\%}$
        & $513.7(1)^{+6.6\%}_{-6.2\%}$
        & $1.46$
        & $528(2)^{+0.2\%}_{-1.5\%}$
        & $1.03$
     \\
    $\PW^-$ pol.\ sum.\
        & $352.91(2)^{+11.5\%}_{-9.3\%}$
        & $512.3(1)^{+6.6\%}_{-6.2\%}$
        & $1.45$
        & $525(1)^{+0.2\%}_{-1.3\%}$
        & $1.03$
     \\
     \hline
  \end{tabular}
  }
  \caption{Total cross sections of $\ppmnj$ for the \emph{fiducial} setup
        at the LHC with a centre of mass of $\sqrt{s}=13\TeV$.  The unpolarised
        and polarised-sum predictions are provided at LO, NLO, and NNLO
        accuracy as well as with the corresponding $K$-factors.  All cross
        sections are expressed in $\fb$.  The numbers in parenthesis indicate the
        numerical error while the ones in per cent represent 7-scale variation.}
  \label{tab:fiducial_xsec_approx}
\end{table}

\begin{figure}[!t]
  \centering
\includegraphics[width=.49\linewidth]{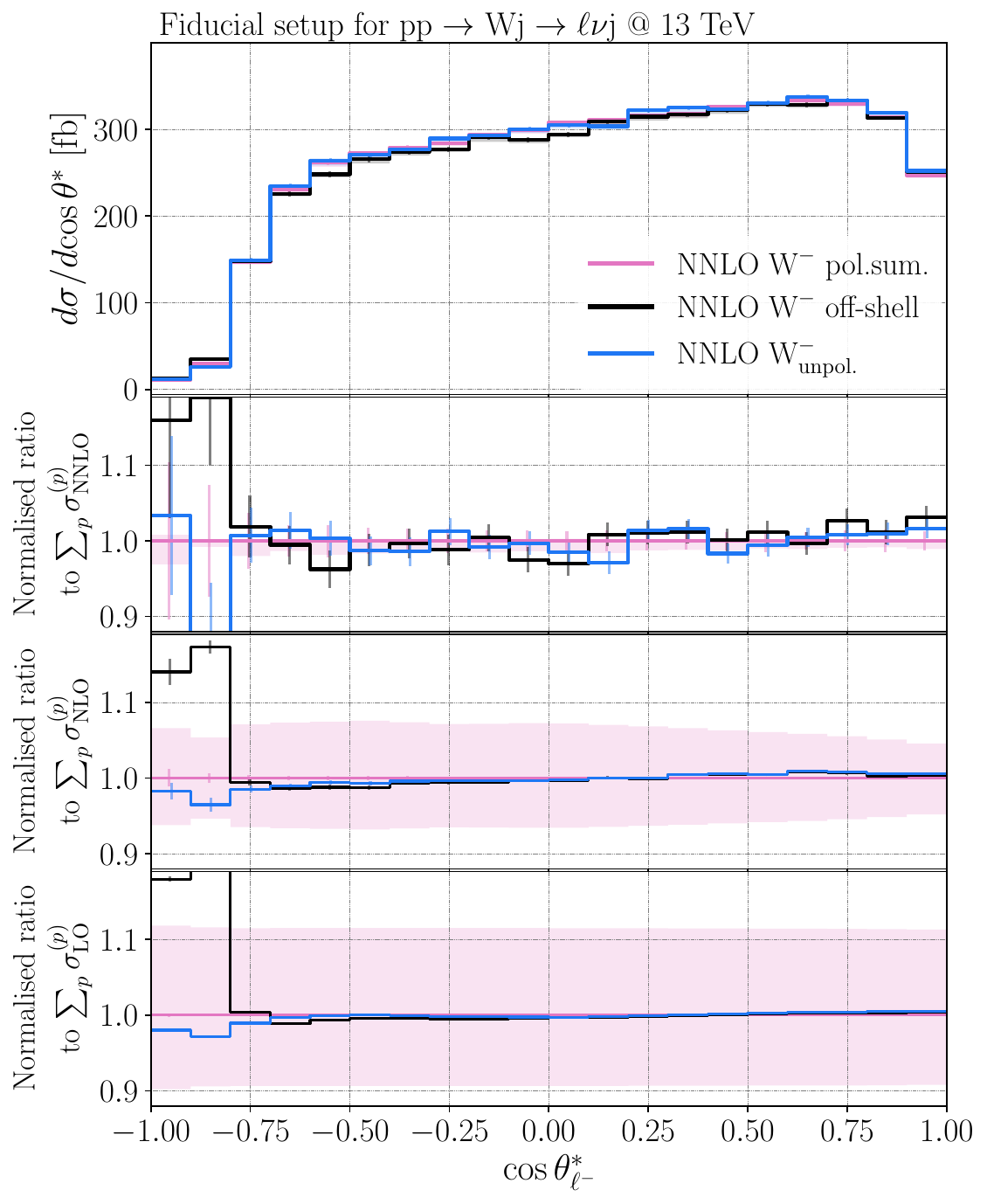}
\includegraphics[width=.49\linewidth]{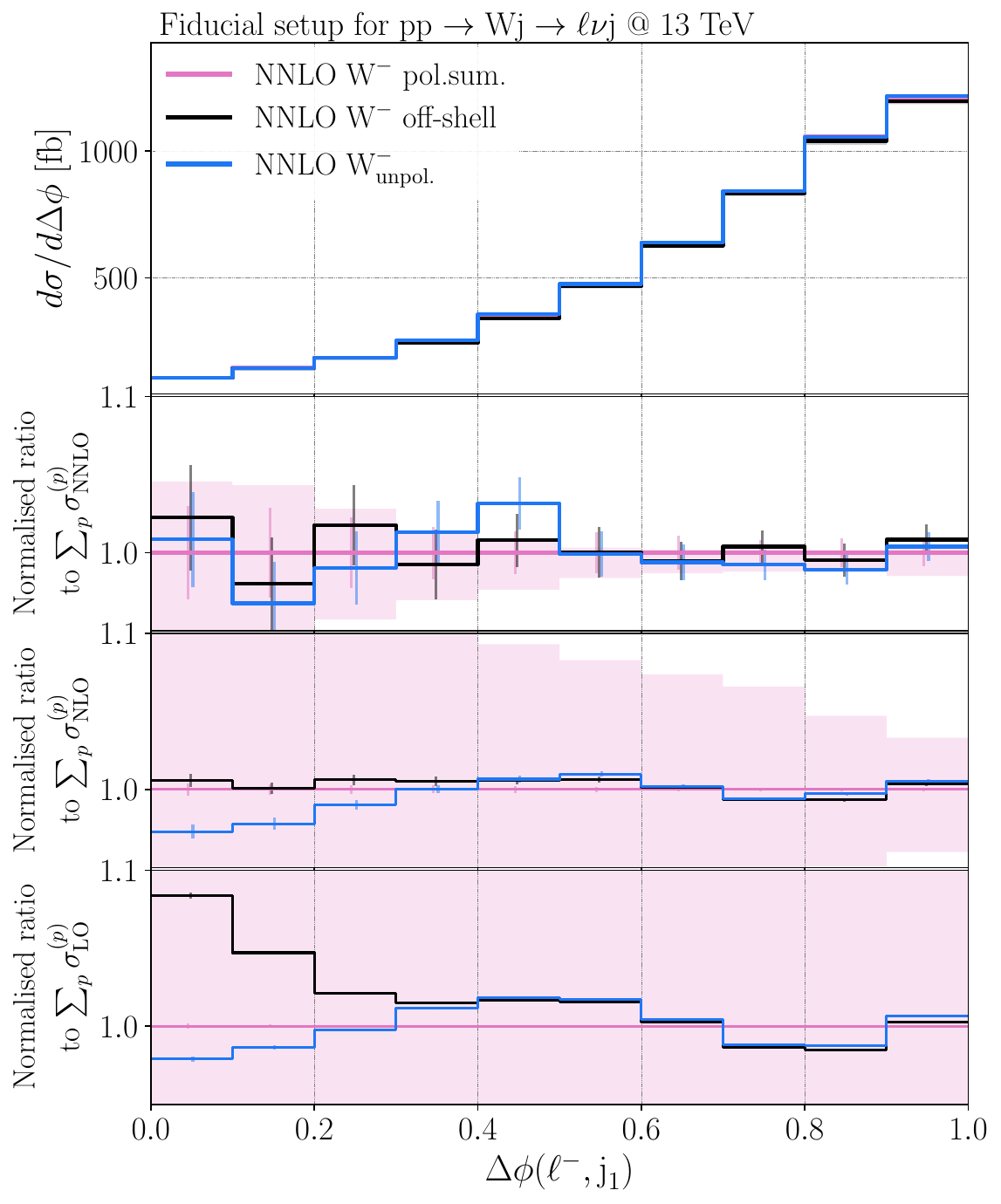}
\\
\vspace{1em}
\includegraphics[width=.49\linewidth]{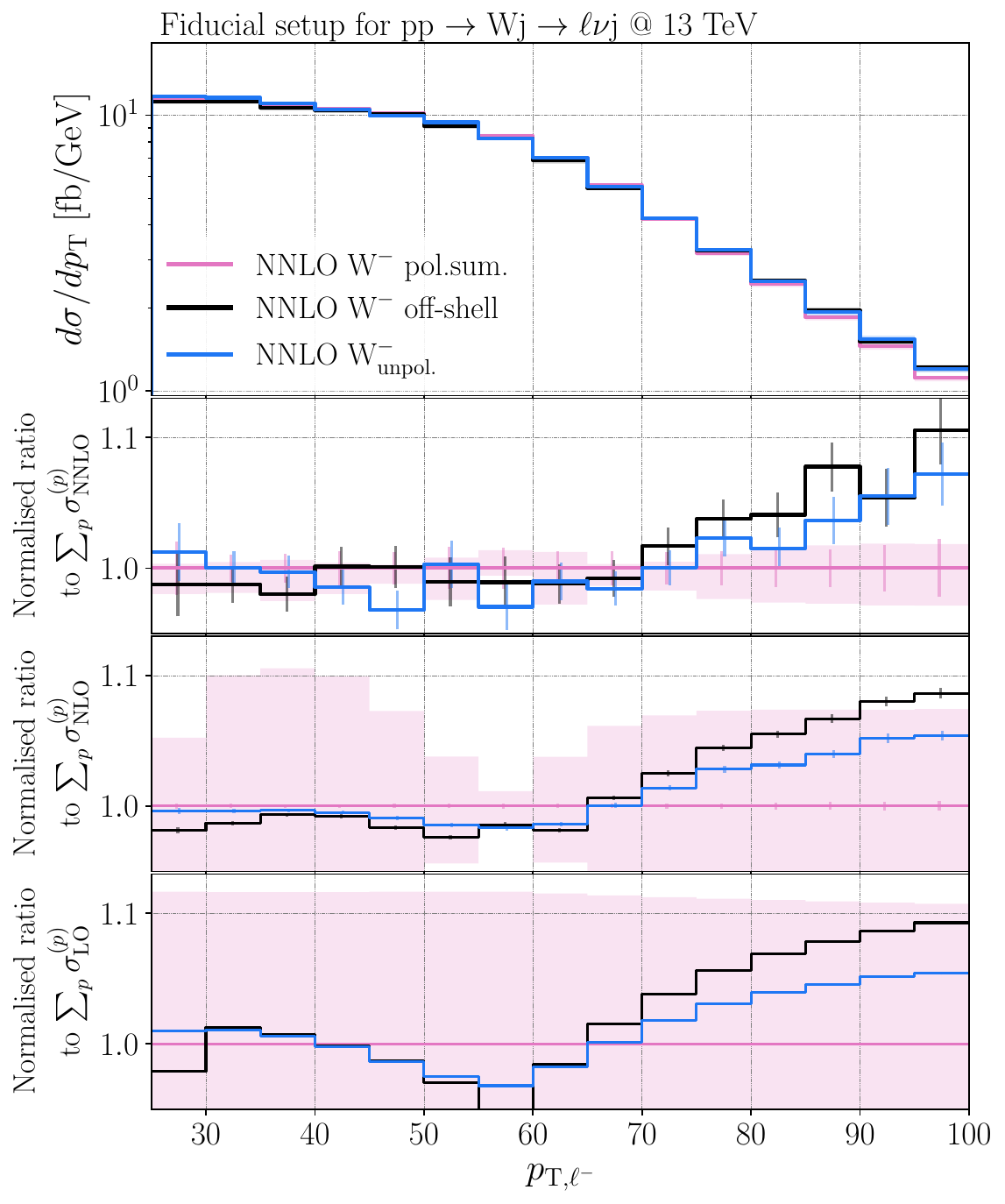}
\includegraphics[width=.49\linewidth]{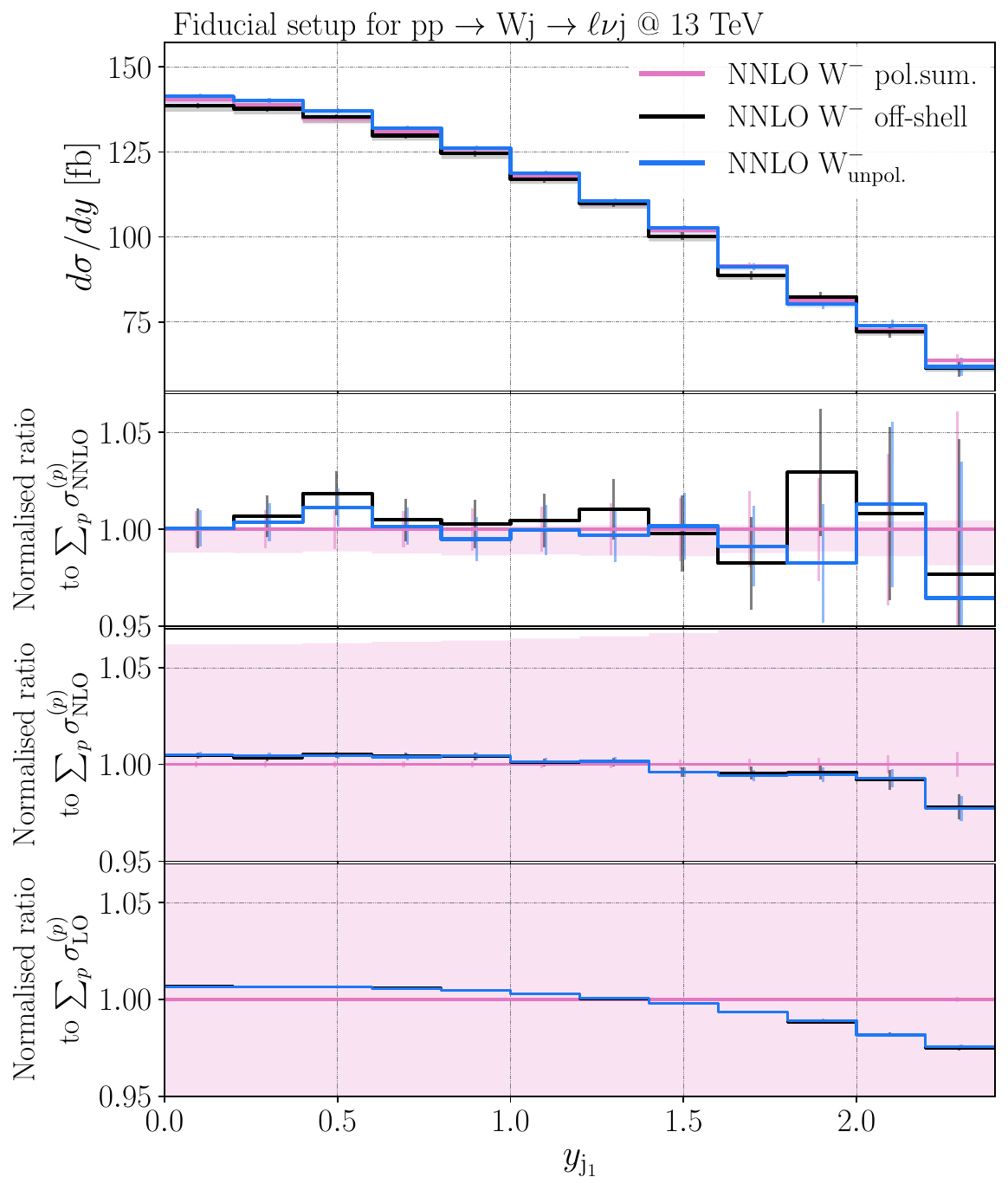}
  \caption{Differential distributions for $\ppmnjM$ in the \emph{fiducial}
    setup: $\cos \theta^*_{\ell^-}$ (top left), $\Delta \phi (\ell^-, \Pj_1)$ (top
    right), $p_{\rm \PW^-}$ (bottom left), and $p_{\rm \ell^-}$ (bottom right).
    The upper pane shows absolute distributions at NNLO QCD in the following setups:
    unpolarised NWA, the sum of polarised NWA, and off-shell.
    The other three panes feature the
    same distributions normalised to the sum of polarised setups at LO, NLO and NNLO.
    The pink band represents
    the 7-point scale variation for the ratio denominator, while the bars indicate
    the Monte Carlo uncertainty.
      \label{fig:polarisation_effects}}
\end{figure}

We turn to differential distributions presented in
\reffi{fig:polarisation_effects}. Starting with the $\cos \theta^*_{\ell^-}$
observable, one can see the
off-shell and polarisation interference effects consistent with zero,
apart from the low angles, where $\cos\theta^*_{\ell} < -0.75$.
This region corresponds to the backwards emission of the charged lepton,
and it is heavily influenced by the phase-space cuts.
Given the distinct polarisation
shapes as presented in \reffi{fig:polarisations_cosTheta} and the absence of
approximation effects elsewhere,
this observable would be perfectly suited to separate polarisations. However,
neutrino momentum is not fully available at experiments, so proxies such
as $L_P$ or $\cos\theta_\text{2D}$ are used.

Looking at the distribution of the azimuthal angle between the charged lepton
and the jet, we see the appearance of some non-trivial interference and off-shell effects.
Interference effects can appear only when the angular integration is restricted,
which is what happens with this observable due to its definition.
The effects are still within scale uncertainty at NNLO.
Note how the off-shell effects reach almost 10\% at LO at lower angles and
vanish at NLO. This part of the phase space at LO corresponds to the charged lepton
and the jet both recoiling against the neutrino --- an unlikely event, with
a higher probability for off-shell W-bosons.
At NLO, due to real radiation, such a topology is no more suppressed,
and thus, this part of the phase space receives a
significant correction which washes out the
off-shell effects at lower order.

Next, we discuss the transverse momentum of the charged lepton.
The plot shows that at $25 < p_\rT < 70 \GeV$ interference and off-shell
effects are rather low, under 3\%, and afterwards start to rise up to
10\% for off-shell and 5\% for interference effects, peaking at
$p_{\rT,\ell} = 110 \GeV$. Given that the longitudinal
polarisation fraction falls off rapidly at higher energies, we do
not investigate the high-energy region. One can see that the magnitude of
the approximation effects drops down for NLO corrections and is consistent
with its level at NNLO.
In conjunction with the distinct polarisation-shape profile,
this observable is
well suited for polarisation studies in the specified region.
The same conclusions apply to the other process signature.

\begin{figure}[!b]
  \centering
\includegraphics[width=.49\linewidth]{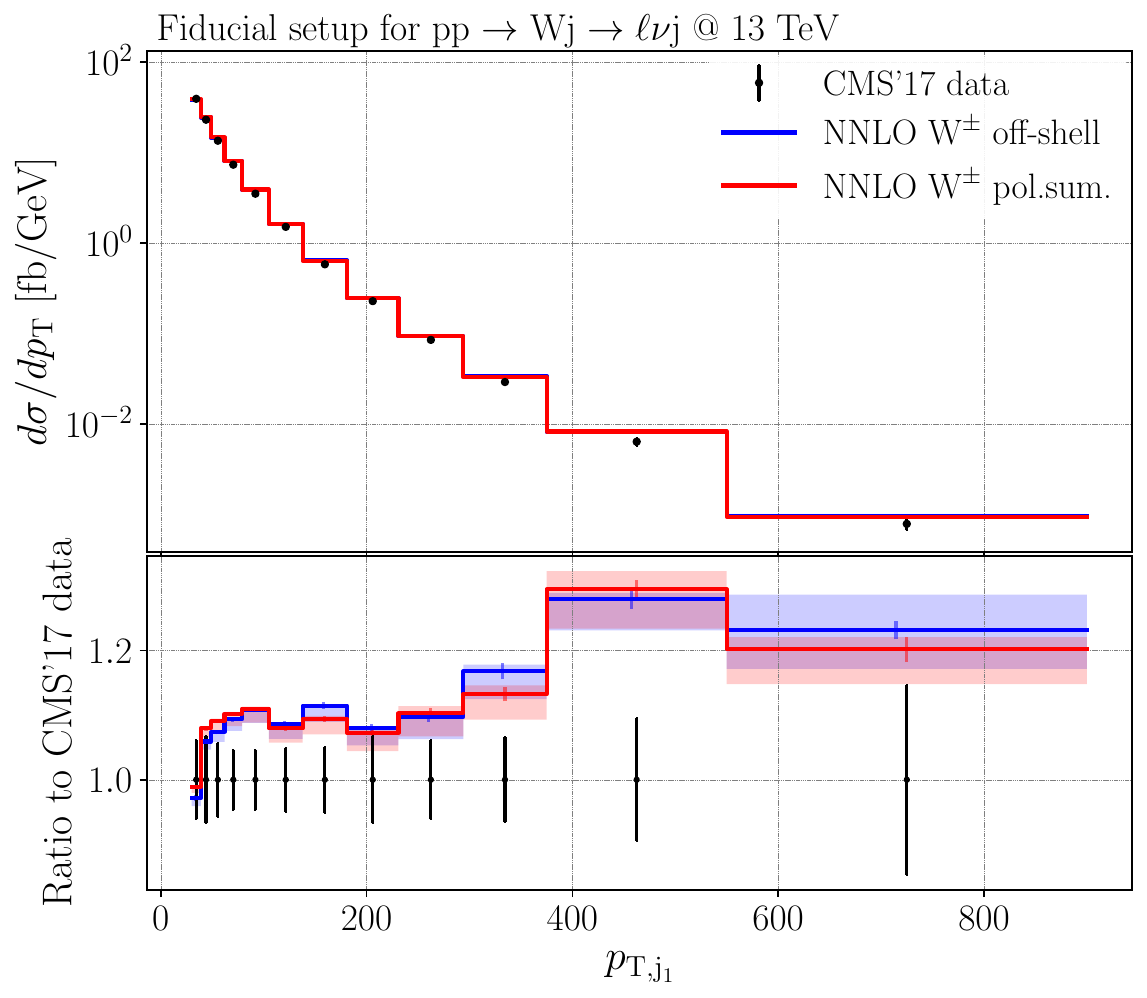}
\includegraphics[width=.49\linewidth]{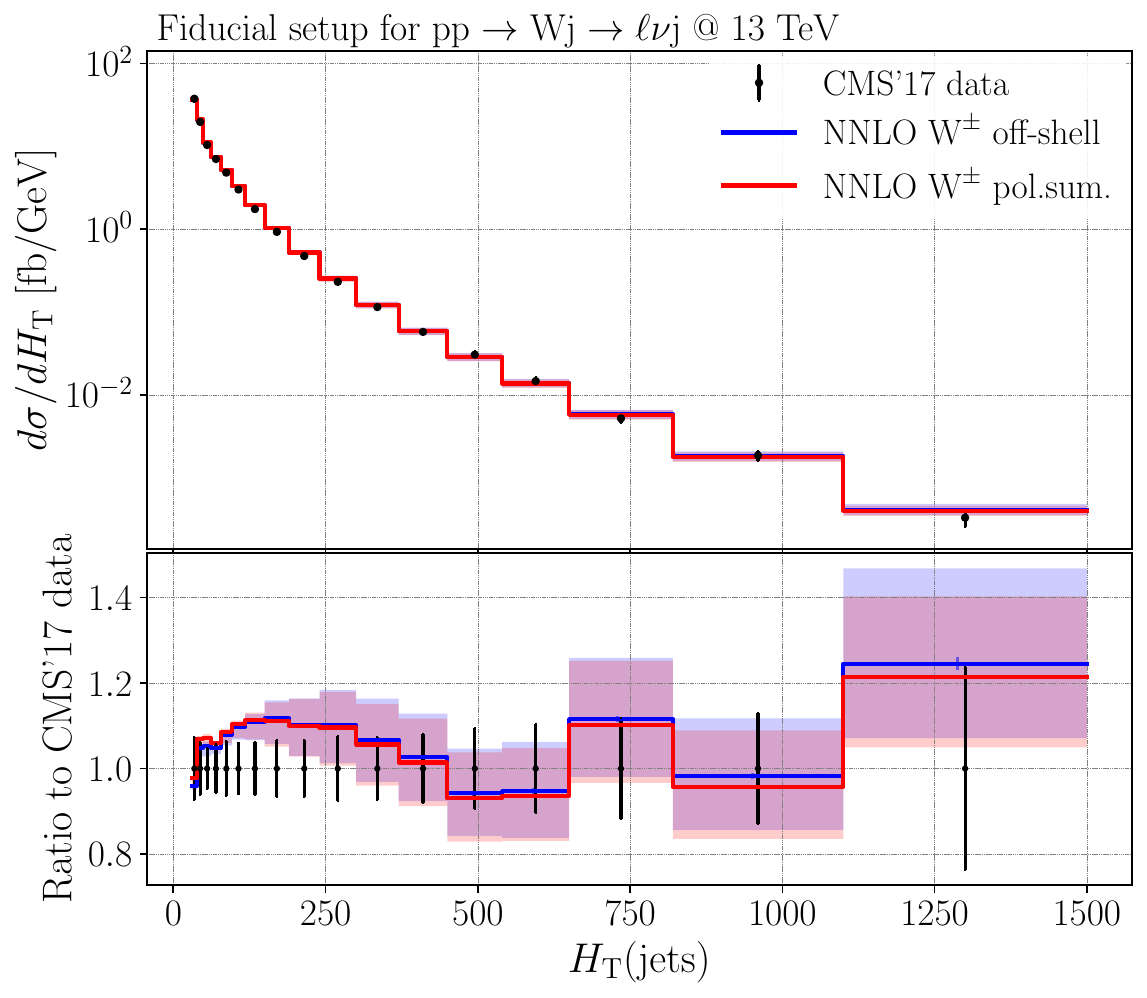}\\
\vspace{1em}
\includegraphics[width=.49\linewidth]{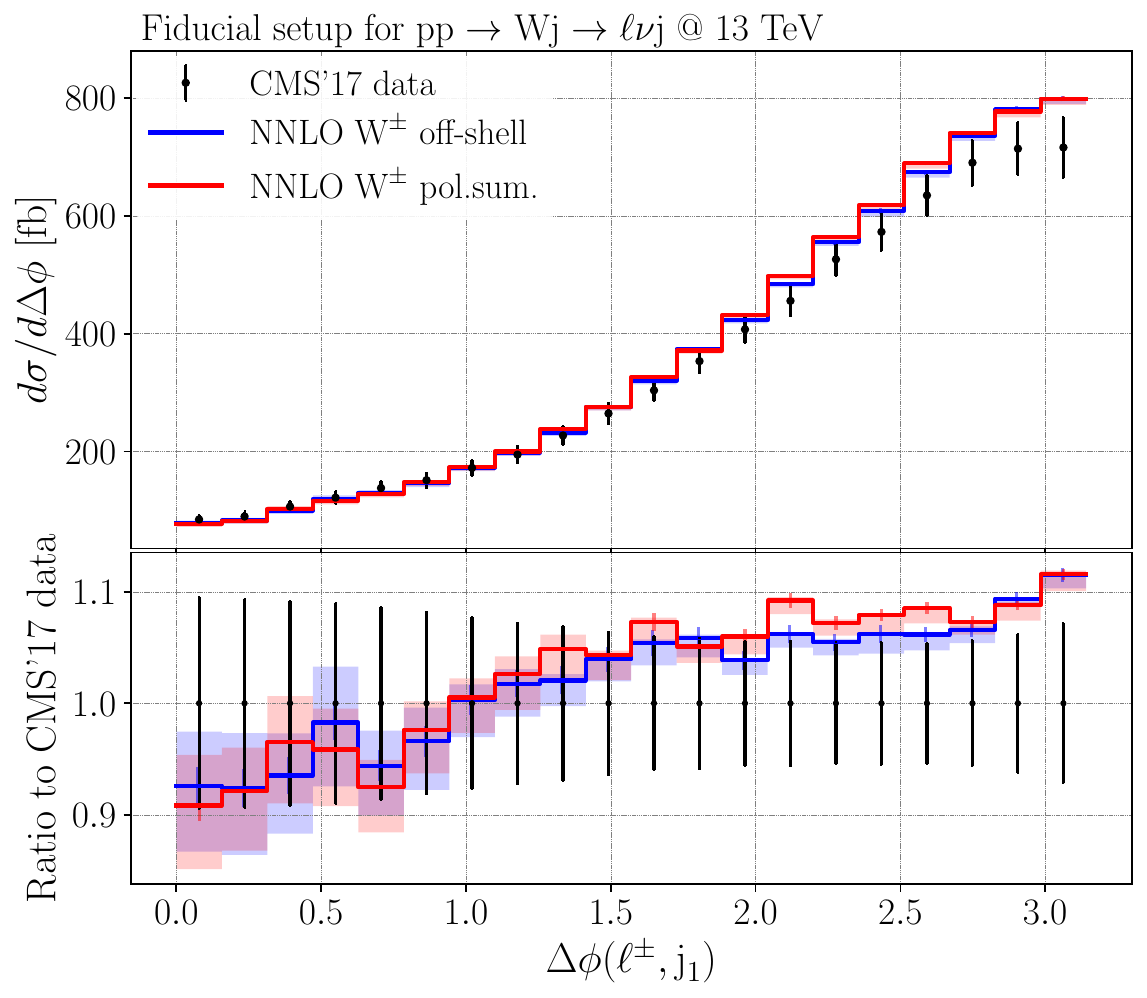}
\includegraphics[width=.49\linewidth]{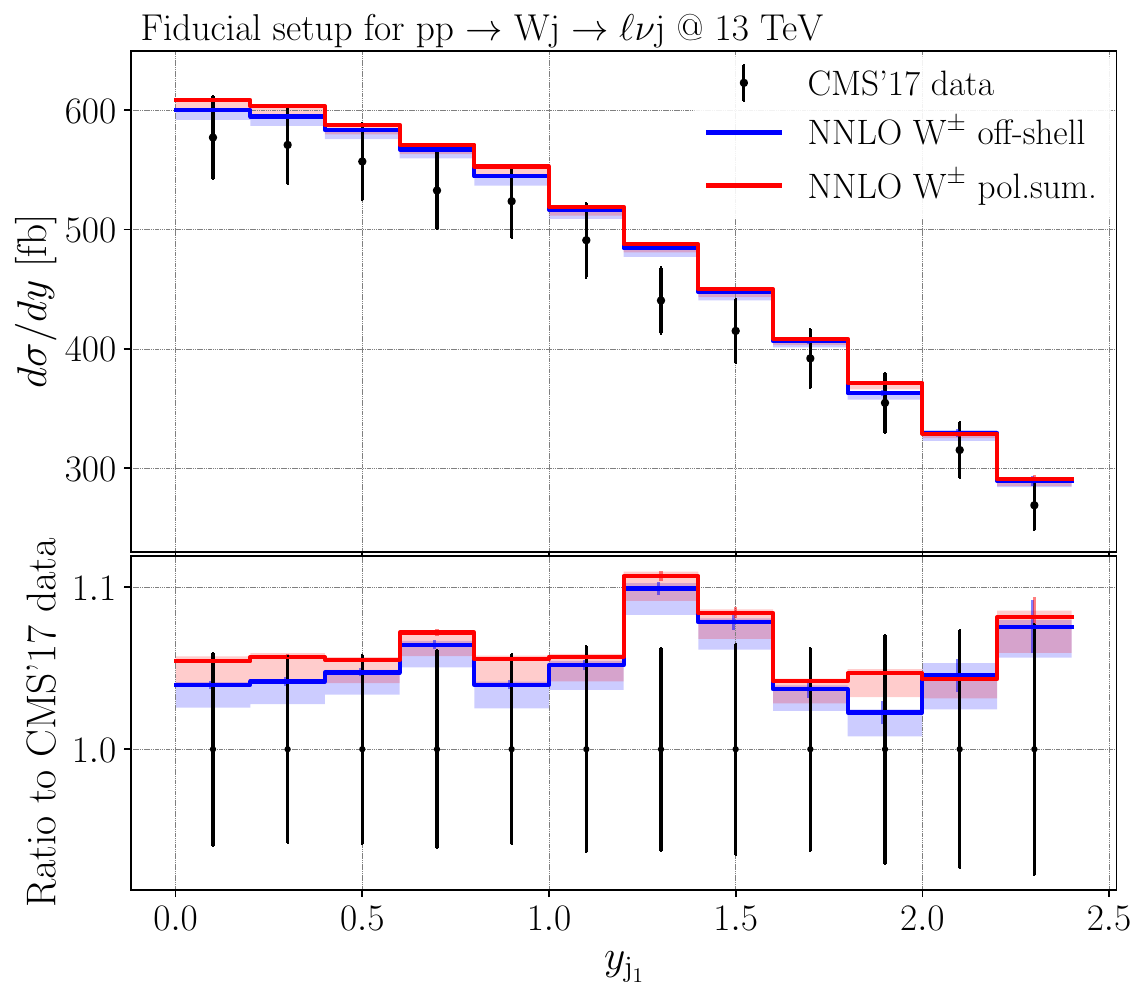}
  \caption{Differential distributions for $\ppmnj$ in the \emph{fiducial} setup:
    transverse momentum of the hardest jet (top left),
    transverse energy of jets (top right),
    azimuthal angle between the charged lepton and the hardest jet (bottom left),
    and rapidity of the hardest jet (bottom right).
    The upper pane features absolute predictions at NNLO
    for the polarised sum in the NWA and the off-shell setup.
    The data points are taken from \citere{CMS:2017gbl}.
    The lower pane shows the same predictions
    normalised to the experimental data. The bands represent
    scale variation uncertainty.
      \label{fig:th_data}}
\end{figure}

Having investigated both off-shell and interference effects differentially,
we can put them in perspective with experimental data. Such an analysis
for several differential distributions is presented in \reffi{fig:th_data}.
Specifically, the sum of polarised predictions and the off-shell
calculation are compared against data for the combined signatures of the W+j process.

In general the agreement is rather good: the data and the
theoretical predictions agree within uncertainties. However,
the NNLO-precise calculations tend to overshoot the data,
both at the level of differential distributions and total cross section.
Missing EW corrections can potentially account for this
effect. They are expected to be negative and have a magnitude
of 5-20\% depending on the energy scale \cite{Denner:2009gj}.
In addition, the data were originally compared with another NNLO calculation
presented in \citere{CMS:2017gbl}, and as an important cross check,
we observed that in all presented distributions, our results are
in agreement at the per-cent level with the exception of $\Delta \phi
(\ell^\pm, \Pj_1)$ at smaller angles, where our result gets to about one scale
band lower (roughly $10\%$),
while still being in agreement with experimental data.
This mismatch originates from the different scale choice. In
ref.~\cite{Boughezal:2016dtm}, the central scale is
$\mu = \sqrt{M^2_{\ell\nu}+\sum_i p^2_{\rm T, \Pj_i}}$.
It coincides with our definition in \refeq{eq:mu} asymptotically at high $p_\rT$,
but behaves differently at low $p_\rT$.
The small angle between the jet and the charged lepton corresponds
to the smaller energies, hence the difference between the two NNLO calculations.

We would like to stress that the shortcomings of the polarised NWA predictions
are negligible in comparison to experimental uncertainties and are mostly
within estimated missing higher-order contributions.
This gives reassurance that boson polarisation studies are well justified
for W+j process and would benefit from more precise experimental measurements.

\subsection{Fitting polarised fractions}
\label{sub:fits_to_polarised_shapes}

In \refse{sec:def} we discussed how to extract longitudinal
and transverse boson polarisations.
Polarisation fractions are simply theoretical quantities that depend on the
intricate structure of the EW sector in the particle theory of the nature.
It means that such parameters can only be experimentally extracted in an
indirect way, based on theoretical inputs. One of the commonly used methods is
template fitting. It prescribes fitting predefined shapes of polarised
(theoretical) predictions to the data. In this section we will show
how such method benefits from NNLO-accurate predictions.

As a first step towards experimental application,
we use \emph{mock data} for which we simply take the NNLO fully
off-shell prediction. The underlying assumption is that these distributions
are best suited to describe the data and can therefore serve as a proxy for a
more precise experimental measurement.
In particular, the mock data errors are generated
according to the NNLO fully off-shell prediction based on a total luminosity of
250 fb$^{-1}$. The fit only considers bins where the off-shell
and interference effects are small: in particular, we use a 3\% criterion.
The polarisation fractions are fitted as linear coefficients that multiply the
normalised polarisation shapes. The more distinct the shapes are, the more
precise the fit will be.
We have also checked that fits to unpolarised and sum of polarised setups
produce similar results.

Both polarisations are fitted independently using a $\chi^2$ fit for each of the 7
scales taking into account all errors:
\begin{equation}
 \chi^2 = \sum\limits_{i\in\text{bins}}\frac{\left(\sigma^i_{\text{off-shell}}
             - \sum\limits_{j=\text{L,T}}\alpha_j\cdot\sigma^i_{\text{pol},j}\right)^2}
             {(\epsilon^i_\text{off-shell})^2
             + \sum\limits_{j=\text{L,T}} (\alpha_j\cdot \epsilon^i_{\text{pol},j})^2},
\end{equation}
where $\eps^i_\text{off-shell} = \sqrt{\sigma^i_{\text{off-shell}} / L}$
is the luminosity-projected error on the off-shell differential distribution,
$\eps^i_{\text{pol},j}$ are the Monte-Carlo errors of the polarised calculation $j$,
and $\alpha_j$ are the fitting coefficients.
We also combine distributions of both signatures,
treating polarisation fractions as charge-independent.
This choice is based on the extreme proximity of the corresponding fractions in
\refta{tab:fiducial_xsec}.
For each of the scale choices, we obtain $1\sigma$ and
$2\sigma$ confidence regions in the $f_\text{L}-f_\text{T}$ plane.
The envelop of all such regions then serves as a proxy for theoretical
uncertainty of the result.
NLO and NNLO predictions differ not only in the size of the scale uncertainty,
but also in the variation of the normalised distribution shapes, which affects
the spread of fits at different scale choices.
To compare the two, we present fits performed using NLO predictions on the left
and NNLO predictions on the right.
The $1\sigma$ and $2\sigma$ confidence regions maintain their form which is
defined by the errors of the data that is being fit, but the spread of the fits changes.
To show how the result of the fit compares with our results from polarised cross section
calculations, we present the error bar corresponding to
polarisation fractions in \refta{tab:fiducial_xsec} and its
theoretical uncertainty.

\begin{figure}[!t]
  \centering
  \includegraphics[width=.49\linewidth]{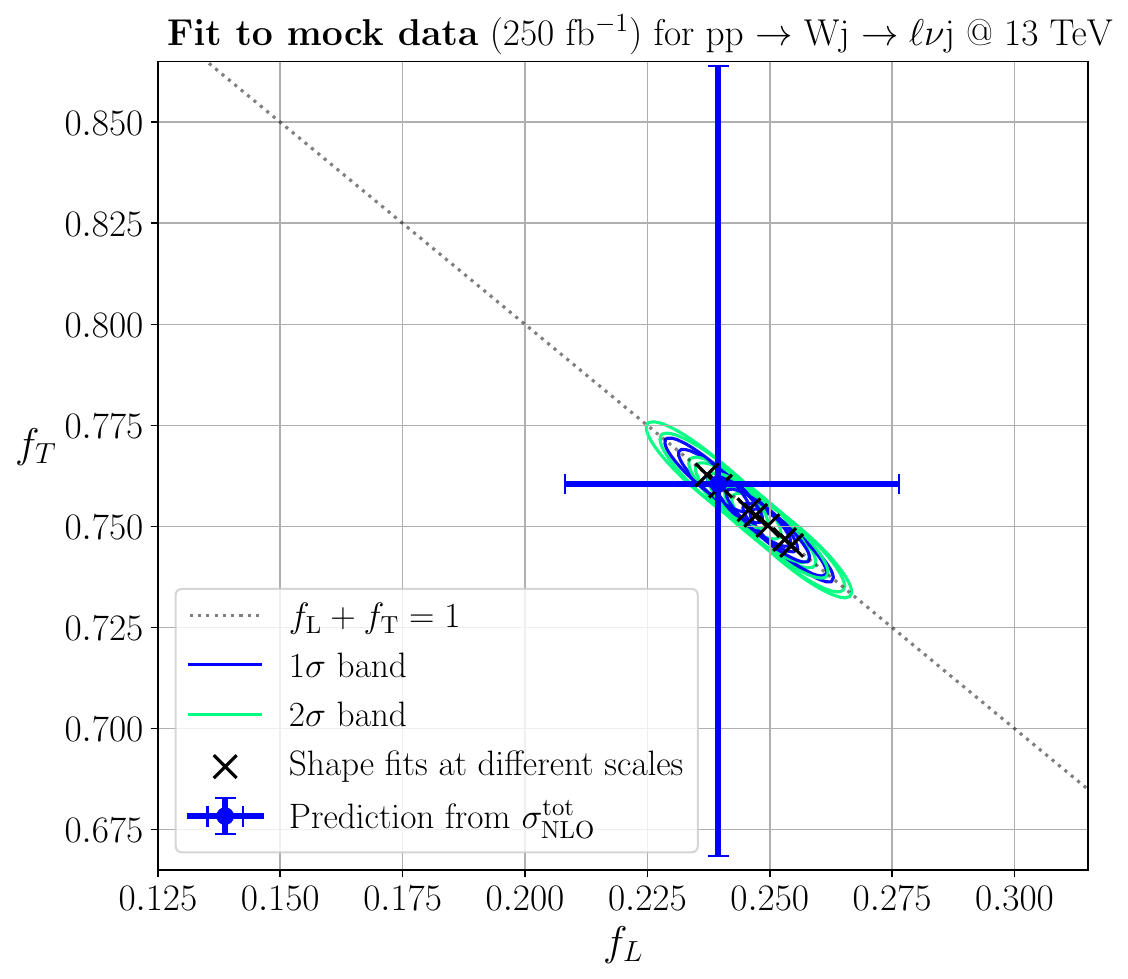}
  \includegraphics[width=.49\linewidth]{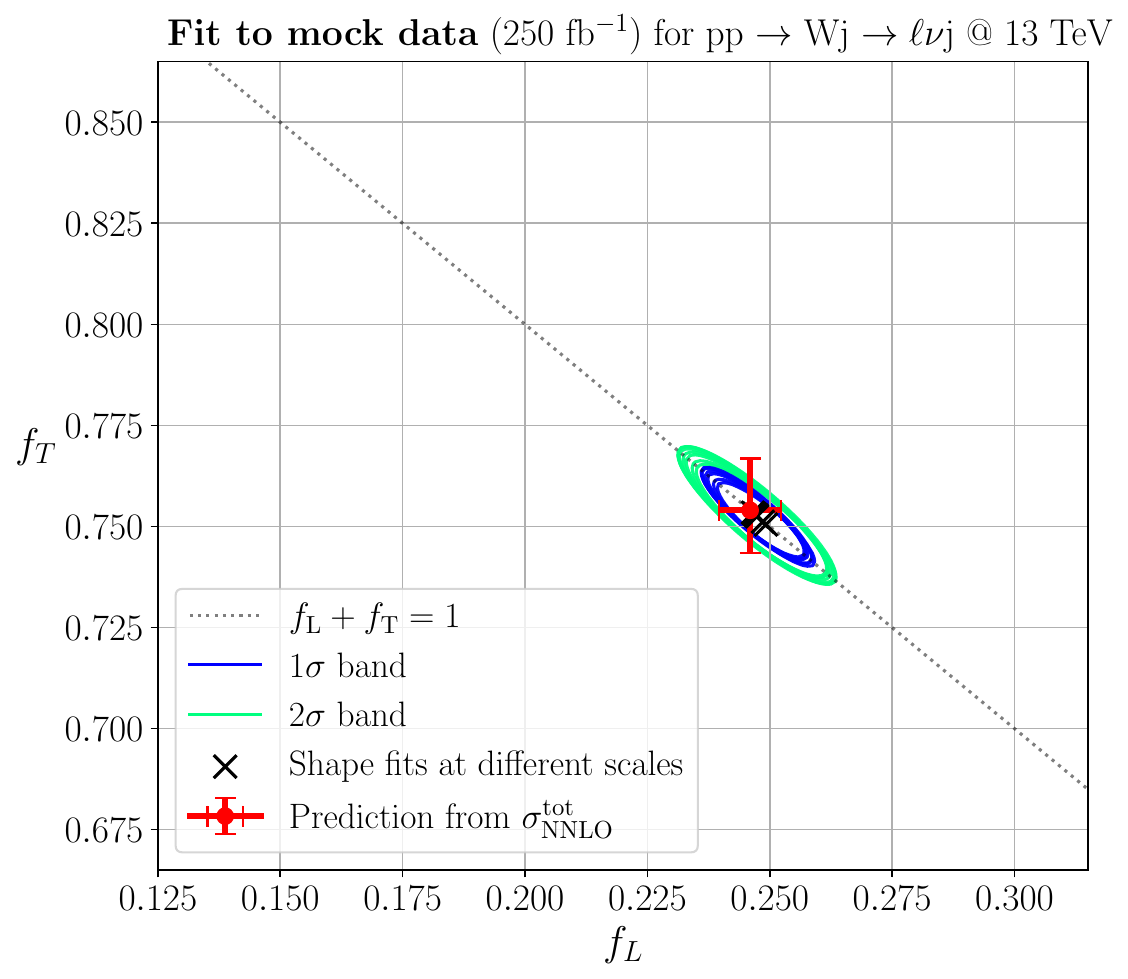}
  \caption{
      Fit to the mock data of the polarisation fractions for both signatures
      $\ppmnj$ in the fiducial setup using
      polar angle of the charged lepton emission
      at NLO (left) and NNLO (right).
      The blue and light green contours
      identify $1\sigma$ and $2\sigma$ regions, respectively.
      Each ellipsis in
      the $f_{\rm T}$ and $f_{\rm T}$ plane
      corresponds to the uncertainty of the fit of one setup in the 7-scale set.
      The error bar represents the prediction for the polarisation fractions
      from \refta{tab:fiducial_xsec}.
      \label{fig:fit_cosTheta}
    }
  \vspace{2em}
  \includegraphics[width=.49\linewidth]{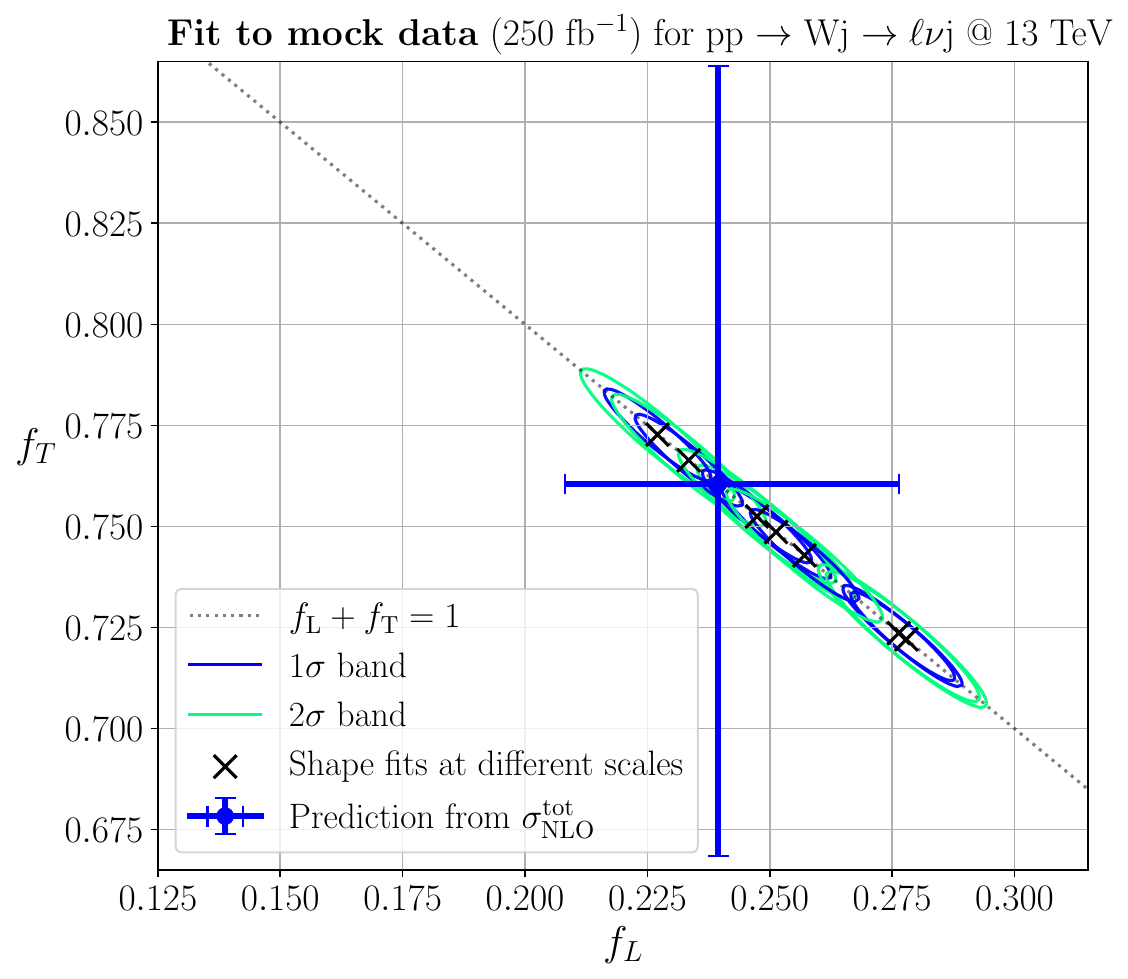}
  \includegraphics[width=.49\linewidth]{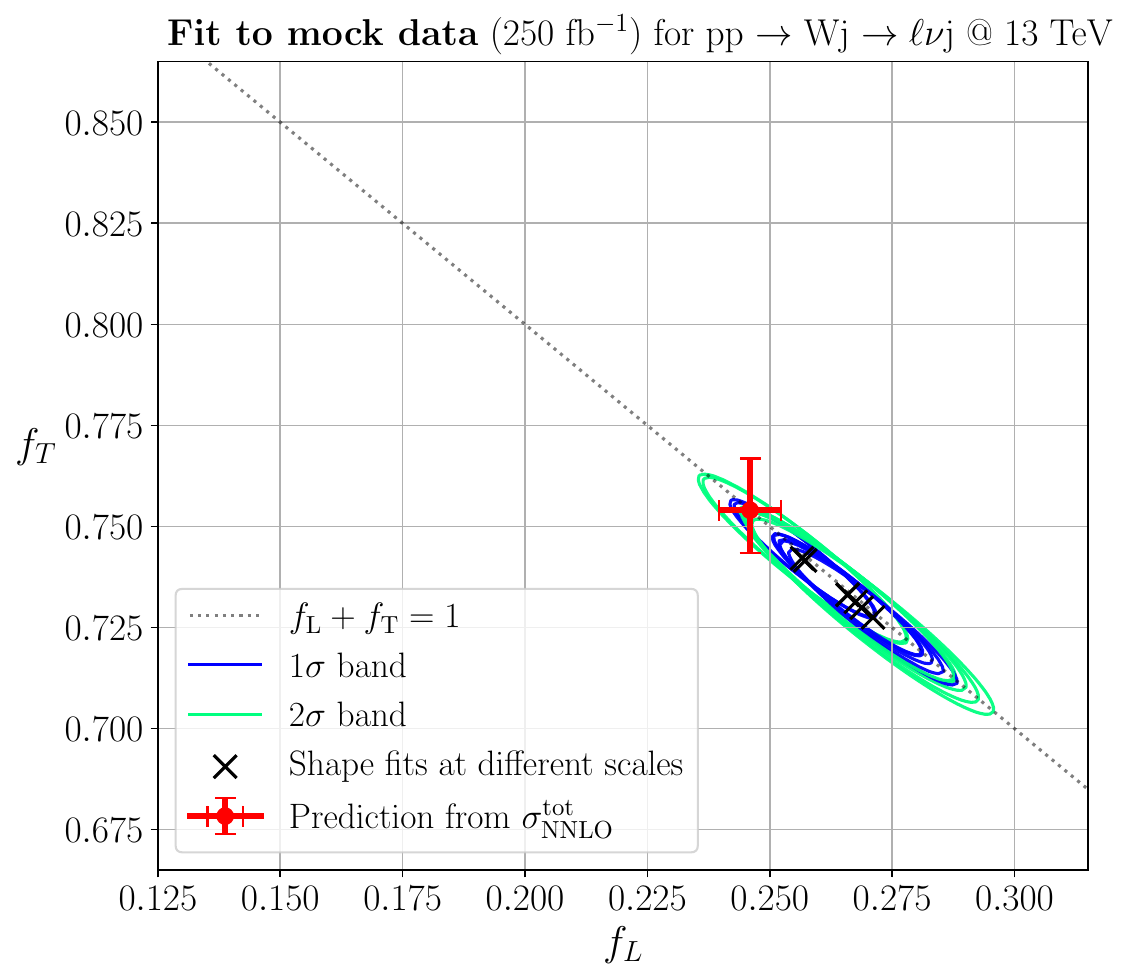}
  \caption{
      Fit to the mock data of the polarisation fractions for both signatures
      $\ppmnj$ in the fiducial setup using
      polar angle between the charged lepton and the hardest jet
      at NLO (left) and NNLO (right).
      Same plot structure as in \reffi{fig:fit_cosTheta}.
      \label{fig:fit_cosTlj1}
    }
\end{figure}

The fit using the polar emission angle of the charged lepton is presented in
\reffi{fig:fit_cosTheta}. The uncertainty on the longitudinal fraction
estimated from the fit shrinks by 50\% at NNLO and is determined
by the size of the ellipse, while the shape variation is drastically reduced.
This observable has the highest power to discern polarisations and thus yields
the lowest uncertainty.
As we observed, the shape variation is generally greatly reduced at NNLO,
in comparison to NLO, but interestingly, the LO shape variation is also very small.
Its fit, however, would be completely meaningless given the large $K$-factors.

Next, we present another observable which is rather sensitive to polarisation
fractions: the polar angle
between the charged lepton and the hardest jet in \reffi{fig:fit_cosTlj1}. The
spread is greatly reduced but not quite as fully as in the previous case. With
data uncertainties taken into account, NNLO corrections reduce the size of the
overall envelope uncertainty by 50\%. This is an excellent
observable which could be used in experimental analyses.

\begin{figure}[!t]
  \centering
  \includegraphics[width=.49\linewidth]{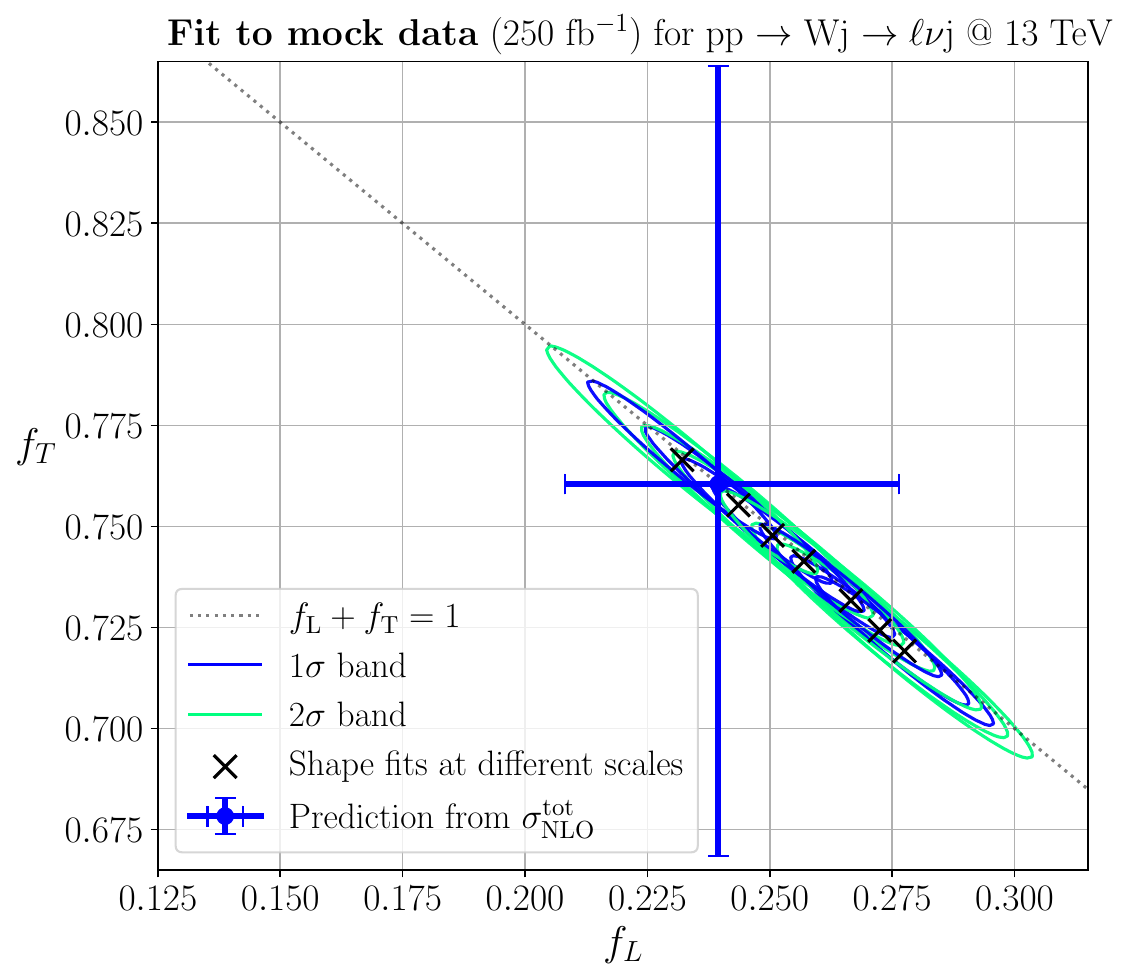}
  \includegraphics[width=.49\linewidth]{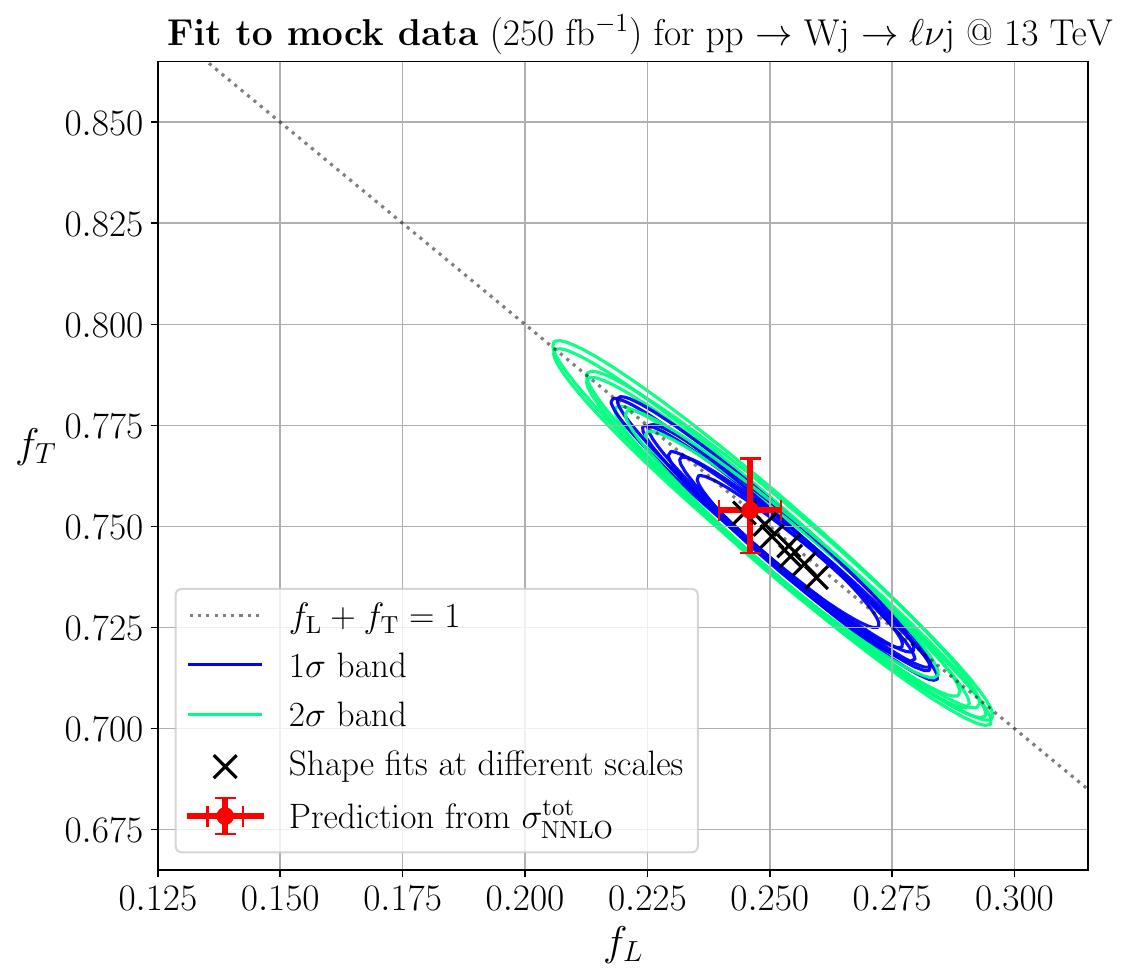}
  \caption{
      Fit to the mock data of the polarisation fractions for both signatures
      of $\ppmnj$ in the fiducial setup using
      the 2D distribution of rapidity and transverse momentum of
      the charged lepton
      at NLO (left) and NNLO (right).
      Same plot structure as in \reffi{fig:fit_cosTheta}.
      \label{fig:fit_pTl_yl}
    }
  \vspace{2em}
    \includegraphics[width=.49\linewidth]{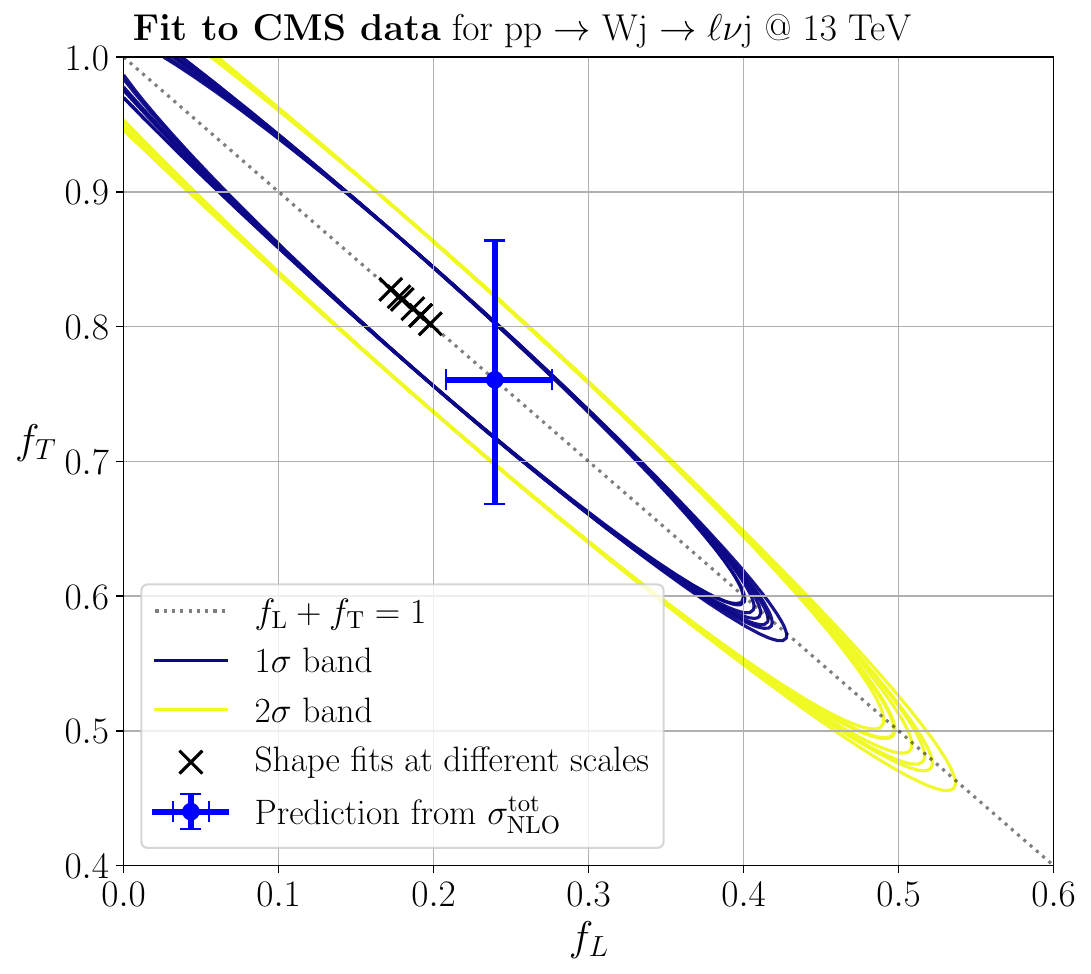}
    \includegraphics[width=.49\linewidth]{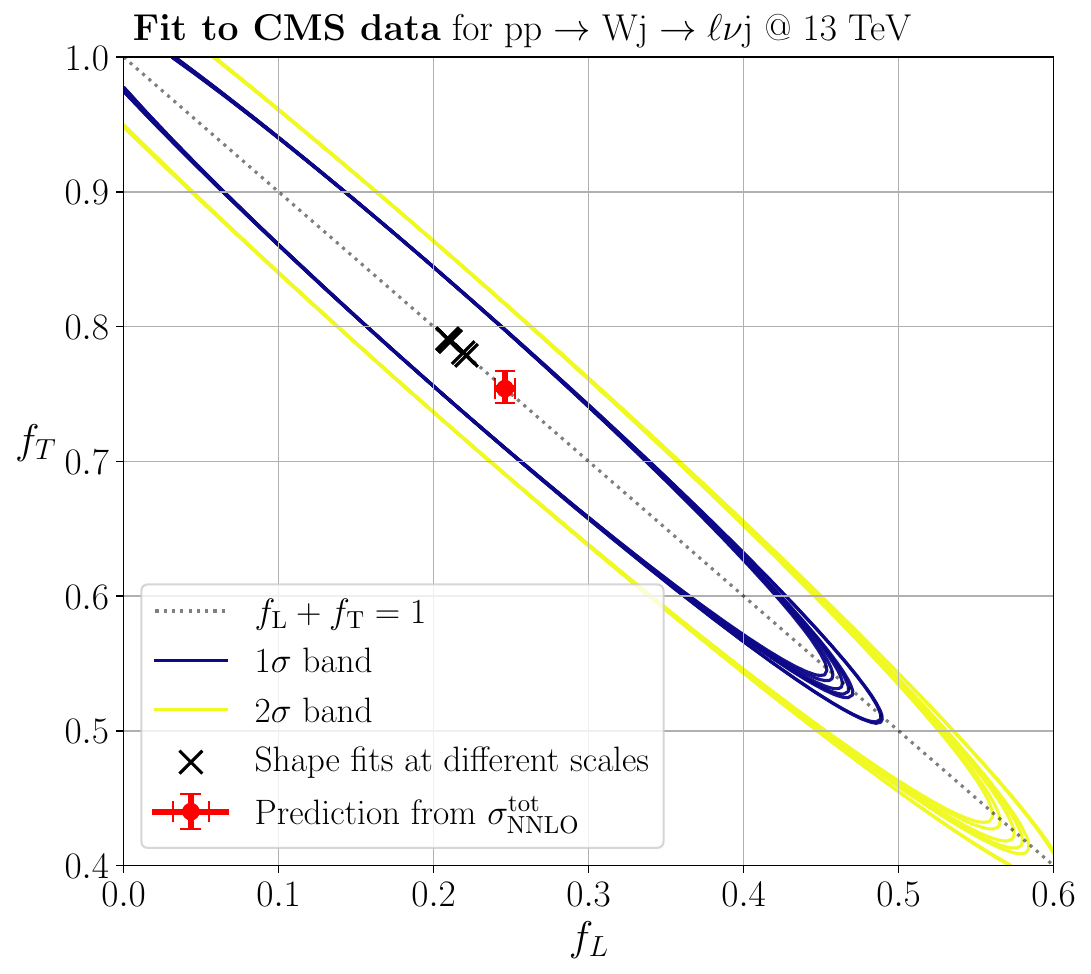}
  \caption{
      Fit to the CMS data \cite{CMS:2017gbl}
      of the polarisation fractions for both signatures $\ppmnj$ in the fiducial setup using
      the $y_{\Pj_1}$ distribution
      at NLO (left) and NNLO (right).
      Experimental systematic errors were separated from statistical errors.
      Statistical errors were further decreased by a factor of 3.
      Same plot structure as in \reffi{fig:fit_cosTheta},
      except for the $2\sigma$ band which is displayed in yellow.
      \label{fig:fitdata_yj1}
    }
\end{figure}

In \reffi{fig:fit_pTl_yl}, we fit to the two-dimensional distribution of the
transverse momentum and rapidity of the charged lepton. After reducing the data
set in the transverse-momentum observable to minimise off-shell and interference
effects, this observable possesses a strong discrimination power.
As mentioned previously, we expect that the pair of $y_{\text{j}_1}$ and
the lepton $p_\text{T}$ will have similar or even stronger polarisation
separation power, since in contrast to lepton rapidity, jet rapidity is
sensitive to polarisation shapes. Still, the evolution of $y_\ell$ shape
at different energies seems to be non-trivial enough to help the presented
2D distribution to perform better than $p_{\text T, \ell}$ alone.

Finally, in \reffi{fig:fitdata_yj1},
we exercise a fit to the experimental data in ref.~\cite{CMS:2017gbl}.
Out of the presented observables, we picked $y_{\text{j}_1}$ as it provides
the strongest sensitivity to polarisations.
Unfortunately, the size of the error bars in ref.\cite{CMS:2017gbl}, which are
based on the measurement with a total luminosity of 2.2 fb$^{-1}$, does not
allow for a sensible fit, suggesting that the longitudinal polarisation fraction
is anywhere in the interval $[0, 0.5]$. This accuracy is expected to
significantly improve with larger statistics during the high-luminosity phase
of the LHC \cite{Azzi:2019yne} and resemble the results obtained in our mock
data fit shown above.

The presented examples here aim at illustrating the role of NNLO corrections in
the polarisation studies. There are many other observables that could be
helpful in discerning
polarisations, \eg $\Delta R$, $L_p$, $\cos\theta^*_\text{2D}$; but as a rule
the fitting results are universally improved, due to the reduction in shape
variation at NNLO.
Combination of independent observables might also improve the fitting
procedure as they can provide extra (independent) information.

To conclude, we would like to emphasise that we have presented all
theoretical ingredients needed for the experimental extraction of polarised
fractions. A detailed analysis of how the fits should be done and
how results should be presented is left for future work. In particular, there
is still a number of open questions. Should the two signatures be fitted
separately or together? How should theoretical uncertainties be taken into account
in the fit? How should one define the overall uncertainty on the fit of the polarisation
fractions?
We believe that these questions should be addressed in collaboration
with experimental groups.

\section{Conclusion}
\label{sec:conclusions}

The exploration of the intricate structure of the EW sector in the SM is
one of the most exciting aspects of the LHC physics. One way of probing
it is through the extraction of polarisation fractions, which are thought to be
sensitive to potential new-physics effects. This article builds upon this idea
by presenting, for the first time, predictions for the polarised production
of W+j at the LHC up to NNLO QCD accuracy in the SM.

In particular, we have studied in detail the behaviour of the transverse and
longitudinal parts in two setups: \emph{inclusive} and \emph{fiducial}. The first
one contains only some jet requirements, whereas the second one reproduces
the event selection of the CMS measurement \cite{CMS:2017gbl} at $13\TeV$.
Application of cuts
has a large impact on the polarisation fraction meaning that a lot of care needs
to be taken when extrapolating results from a fiducial volume to the inclusive
phase space. Also, similarly to what was shown in other processes,
different polarised processes receive different QCD corrections
up to NNLO accuracy.
This advocates for the use of tailored predictions in experimental analysis.

As the usage of polarised prediction relies on a number of approximations,
we have investigated their effects in detail. In particular, we
studied off-shell effects arising due to NWA and interferences
between polarisation states.
Compared to missing higher-orders estimated through scale variation, such
effects are rather suppressed. In the same way, the experimental uncertainties
are significantly larger than the shifts induced by such approximations.
Overall, we conclude that for $\PW^\pm\Pj$ production at the LHC, the use of polarised
predictions within the NWA is without a doubt a very good approximation of the
full process.

Finally, we have performed global fits of the polarisation fractions using
mock data (the off-shell computations) as well as experimental data. By doing
so, we have singled out the most sensitive observables for the
determination of polarisation fractions. Also, we have shown that with
sufficiently precise data, using NNLO predictions allows for a more precise
determination of polarisation fractions than with lower-order predictions.

In the present work, we have presented all
theoretical ingredients necessary for the experimental extraction of
polarisation fractions at the LHC. In particular, we will happy
to provide all theoretical predictions (cross sections and differential
distributions) in data format upon request. These could be used
right away by experimental collaboration to extract polarisation fractions with
existing data, hence paving the way for the precise exploration of the EW
sector of the SM in the high-luminosity phase of the LHC.

\section*{Acknowledgements}\label{sec:acknowledgements}

The authors would like to particularly thank Micha\l{} Czakon for making
the {\sc Stripper} library available to us and Alexander Mitov for inspiring
discussions. We are also grateful to Lorenzo Tancredi for useful discussions
on the implementation of the polarised two-loop amplitude and Jean-Nicolas Lang
for providing a polarisation-capable private version of {\sc Recola} used for
cross checks.
This research has received funding from the European Research
Council (ERC) under the European Union’s Horizon 2020 Research and Innovation
Programme (grant agreement no. 683211).
M.P. acknowledges support by the German Research Foundation (DFG)
through the Research Training Group RTG2044.
A.P. is also supported by the Cambridge Trust and Trinity College Cambridge.
R.P. acknowledges the support from the Leverhulme Trust and the Isaac Newton Trust,
as well as the use of the DiRAC Cumulus HPC facility under Grant No. PPSP226.

\appendix
\section{Polarised two-loop matrix elements}
\label{sec:appendix}

The two-loop amplitudes for the partonic process
\begin{align}
  \bar{\Pq}' \Pq \to \PW \Pg,
\end{align}
and the corresponding crossed ones, are a necessary input for the
NNLO QCD corrections to the polarised cross sections. The amplitudes for the
off-shell process, including the leptonic decay of the $\PW$-boson,
have been presented
in \citere{Gehrmann:2011ab}. We use their notation in
what follows. The essential result of this work is the independent
helicity structure coefficients $\alpha$, $\beta$, and $\gamma$ --- also
published explicitly up to two-loop order. The amplitude
$\mathcal{M}(\bar{\Pq}'\Pq \to \PW \Pg)$ can be written in terms of the partonic
current $S^{\mu}$ and the polarisation vector $\epsilon_\mu$ of the
$\PW$-boson:
\begin{align}
  \mathcal{M}(q,h) = \epsilon_\mu(q,h)S^{\mu}\;,
\end{align}
where the dependence on the $\PW$-boson momentum and polarisation $h$ has been made
explicit. Furthermore, the right-handed partonic current $S^{\mu}_R$ can be
decomposed into helicity structures as follows:
\begin{align}
S_R^\mu&(p_1^-;p_3^+;p_2^+) = \frac{1}{\sqrt{2}}
\langle 1 2\rangle [1 3]^2
\left( p_{1 \mu} A_{11} + p_{2 \mu} A_{12} + p_{3 \mu} A_{13} \right)
- \frac{1}{\sqrt{2}} \frac{\langle1 2\rangle [1 3]}{\langle 2 3\rangle }
[1\;| \gamma_\mu |\;2 \rangle \; s_{23} B \notag \\
&+ \frac{1}{\sqrt{2}}
[1 3] [3\;| \gamma_\mu|\;2\rangle \;
\left[ s_{23}B + \frac{1}{2} \left( (A_{11}+A_{12})s_{12}
 + (A_{11}+A_{13})s_{13} + (A_{12}+A_{13})s_{23} \right) \right] ,
\end{align}
featuring the scalar coefficients $A_{11}$, $A_{12}$, $A_{13}$, and $B$. After
contraction with the decay current, only three linear combinations of these
coefficients ($A_{11}$, $A_{12}+2B$, and $A_{13}$) remain. These combinations
are linearly related to the already mentioned coefficients $\alpha$, $\beta$,
and $\gamma$, but it worth mentioning that the $A_{1j}$ and $B$ coefficients cannot
be obtained separately from $\alpha$, $\beta$, and $\gamma$. Due to
transversality of the polarisation vector $\eps_\mu$, one expects that this
reduction of linear independent coefficients occurs in the same way for the
on-shell amplitude as well. The purpose of the rest of this note is to
explicitly show this property.

We start by decomposing the polarisation vector in terms of a momentum basis
which spans the four dimensional space:
\begin{align}
 \epsilon^\mu = c_1 p_1^\mu + c_2 p_2^\mu + c_3 p_3^\mu
    + c_4 \epsilon^{\mu}_{p_1,p_2,p_3}/s_{12}\,,
 \label{eq:epsdecomp}
\end{align}
where $\epsilon^{\mu}_{p_1,p_2,p_3} =
\epsilon^{\mu\nu\rho\sigma}p_{1,\nu}p_{2,\rho}p_{3,\sigma}$ is the contracted
Levi-Civita symbol and $c_i$ are scalar coefficients.
With the usage of the standard spinor-helicity identities,
the contraction of $S^{\mu}_R$ with the external momenta
$p_1,p_2,p_3$ and $\epsilon_{p_1,p_2,p_3}$ leads to
\begin{align}
 S^{p_1}_R = p_{1 \mu }S^\mu_R =& -\frac{\langle 1 2\rangle [1 3]^2}{2\sqrt{2}}
      \bigl(A_{11}(s_{12}+s_{13})+A_{13}s_{23}+(A_{12}+2B)s_{23} \bigr) ,\\
 S^{p_2}_R = p_{2 \mu }S^\mu_R =& \frac{ \langle 1 2\rangle [1 3]^2}{2\sqrt{2}}
       \bigl(A_{11} s_{12} + A_{13} s_{23}\bigr) ,\\
 S^{p_3}_R = p_{3 \mu } S^\mu_R =& \frac{ \langle 1 2\rangle [1 3]^2}{2\sqrt{2}}
        \bigl(A_{11} s_{13} + (A_{12}+2B)s_{23} \bigr) ,\\
 S^\epsilon_R = \epsilon_{\mu,p_1,p_2,p_3} S^\mu_R =&
  -\frac{i\langle 1 2\rangle [1 3]^2}{4\sqrt{2}} s_{23}
   \bigl(A_{11}(s_{12}+s_{13}) +(A_{12}+2 B)(s_{12}+s_{23})\nonumber \\
  & +A_{13}(s_{13}+s_{23})\bigr)\,,
\end{align}
where the latter contraction with the Levi-Civita symbol utilised the Chisholm identity
to write
\begin{align}
 \epsilon_{\mu,p_1,p_2,p_3} \langle i | \gamma_{\mu} | j ] =
  i/2 \bigl( [ i | 1] \langle 1 | 2 \rangle [2|3] \langle 3 | j \rangle
 - [ i | 3 ] \langle 3 | 2 \rangle [2|1] \langle 1 | j \rangle \bigr)\;.
\end{align}
The contractions now depend (as expected) only on three linear combinations of
coefficients $A_{11},A_{13}$, and $A_{12}+2B$ which are directly related to the
coefficients $\alpha,\beta$ and $\gamma$. The amplitude can be evaluated for
any given $\epsilon_{\mu}$ by solving eq.~\eqref{eq:epsdecomp} for the
coefficients $c_i$ and in the end reads
\begin{align}
  \epsilon_{\mu}S_R^{\mu} = c_1 S^{p_1}_R + c_2 S^{p_2}_R +
    c_3 S^{p_3}_R + c_4/s_{12} S^{\epsilon}_R\,.
\end{align}

\bibliographystyle{JHEPmod}
\bibliography{refs}

\end{document}